\documentclass[iop]{emulateapj}
\usepackage{amssymb,amsmath}
\usepackage{graphicx}
\usepackage{color}
\usepackage{draftcopy}

\slugcomment{version \today: fm}

\shorttitle{CLaSPS: knowledge extraction from astronomical datasets}
\shortauthors{R. D'Abrusco et al.}

\begin{document}
\title{CLaSPS: a new methodology for knowledge extraction 
from complex astronomical datasets}
\author{R. D'Abrusco\altaffilmark{1}, G. Fabbiano\altaffilmark{1}, G. Djorgovski\altaffilmark{2}, 
	    C. Donalek\altaffilmark{2}, O. Laurino\altaffilmark{1}, G. Longo\altaffilmark{3,2}}
\affil{Harvard-Smithsonian Center for Astrophysics, 60 Garden Street, Cambridge, MA 02138}
\affil{Department of Astronomy, California Institute of Technology, MC 249-17 1200 East California 
Blvd, Pasadena CA 91125}
\affil{Dipartimento di Scienze Fisiche, Universit\'a di Napoli ``Federico II", Via Cinthia I-80126, 
Naples, Italy}

\begin{abstract}

In this paper we present the Clustering-Labels-Score Patterns Spotter (CLaSPS), a new 
methodology for the determination of correlations among astronomical 
observables in complex datasets, based on the application of distinct unsupervised clustering techniques. 
The novelty in CLaSPS is the criterion used for the selection of the optimal clusterings, 
based on a quantitative measure of the degree of correlation between the cluster memberships 
and the distribution of a set of observables, the \emph{labels}, not employed for the clustering.
CLaSPS has been primarily developed as a tool to tackle the challenging complexity of the multi-wavelength 
complex and massive astronomical datasets produced by the federation of the data from modern automated 
astronomical facilities. In this paper we discuss the applications of CLaSPS to 
two simple astronomical datasets, both composed of extragalactic sources with photometric observations at 
different wavelengths from large area surveys. The first dataset, CSC+, is composed of optical quasars
spectroscopically selected in the SDSS data, observed in the X-rays by Chandra 
and with multi-wavelength observations in the 
near-infrared, optical and ultraviolet spectral intervals. One of the results of the application of CLaSPS to 
the CSC+ is the re-identification of a well-known correlation between the 
$\alpha_{\mathrm{OX}}$ parameter and the near ultraviolet color, in a subset of CSC+ sources with 
relatively small values of the near-ultraviolet colors.
The other dataset consists of a sample of blazars for which photometric observations 
in the optical, mid and near infrared are available, complemented for a subset of the sources, by  
Fermi $\gamma$-ray data. The main results of the application of CLaSPS to such datasets 
have been the discovery of a strong correlation between the multi-wavelength color distribution of 
blazars and their optical spectral classification in BL Lacs and Flat Spectrum Radio Quasars (FSRQs)
and a peculiar pattern followed by blazars in the WISE mid-infrared colors space. 
This pattern and its physical interpretation have been discussed in details in other 
papers by one of the authors.
\end{abstract}


\section{Introduction}
\label{sec:introduction}

The advancement of discovery in astronomy, from the statistical point of view, can be described 
as the successful application of several distinct Knowledge Discovery (KD) techniques to increasingly
larger data samples. These techniques include: the classification of sources according to one or more 
observational quantities; 
pattern recognition for the discovery of correlations among observable quantities; outlier selection
for highlighting rare and/or unknown sources; regression, for the estimation of derived empirical 
properties from observed quantities. The discovery of new or unexpected correlations between 
observable quantities at different wavelengths, for example, has propelled the understanding of the 
nature of astronomical sources and their physical modeling (see, for example, the discovery of the 
fundamental plane of elliptical galaxies~\citep{djorgovski1987}), and the discovery of the link of the 
galaxy X-ray emission with different stellar populations~\citep{fabbiano1985,fabbiano2002}.

The effectiveness of pattern recognition techniques for the determination of correlations in low 
dimensional spaces (two or three dimensions) has usually relied on the ability of the astronomers to 
visualize the distribution of data and make informed guesses about the nature of these patterns, based on 
theoretical models, reasonableness and intuition. However, this approach becomes more and more ineffectual 
with the increase in complexity and size of the explored datasets. This difficulty has led to the introduction of 
KD techniques in the astronomical context. Such techniques are based on statistical 
and computational 
methodologies capable of automatically identifying useful correlations among parameters in a N-dimensional 
dataset without any \emph{a priori} assumption on the nature of both data and the sought out patterns. 
Using these techniques, the focus of the astronomer can shift to the definition of the general problem 
to be investigated, the selection of the interesting patterns and their physical interpretation.
In this paper, we present CLaSPS, a new methodology based on KD techniques for the exploration 
of complex and massive astronomical datasets and the detection of correlations among 
observational parameters. While CLaSPS is designed for datasets containing very large number
of sources, it is also well suited to handle small datasets, as will be shown in this paper.  

The adoption of KD methodologies in astronomy has only recently surged, due to the increasing availability 
of massive and complex datasets that would be almost intractable if tackled with the knowledge extractions 
techniques classically employed in astronomical research. A review of the advantages and most 
interesting applications of KD to astronomical problems can be found in \citep{ball2010}. The main reasons 
for the delay in the adoption of such methods in astronomy are: a) datasets for which KD has an 
edge over classical methods (because of their size and complexity) have become frequent only 
in the last $\!\sim$15 years; b) slow transition 
from model-driven to data-driven research; c) lack of interdisciplinary expertise required for the 
application of KD techniques. Other disciplines for which the problem of dealing with massive datasets
arose earlier, instead, have seen a steadier and faster growth of the 
number and importance of the KD tools employed on a regular basis. For example, the study of 
financial markets and complex 
networks and systems (applied to the WWW, advertisement placement, epidemiology, genetics, 
proteomics and security) have been on the forefront of application and development of KD 
techniques. Thorough reviews of the applications of KD methodologies to specific 
financial topics, i.e. customer management and financial fraud detection, can be found 
in~\citep{ngai2009} and~\citep{ngai2011} respectively, while a general review of the role of KD in 
bio-informatics is provided in~\citep{natarajan2005}. Even if a certain degree of inter-disciplinary
expertise is desired, domain-specific knowhow is crucial to narrow down the types and number of 
techniques that can be used to address the specific problems encountered in each field, and to 
interpret correctly the results of the application of such techniques to the data. Furthermore, 
KD is only one of the skills necessary to tackle the new problems arising with the onset of 
data-driven astronomy, the other being astrostatistics~\citep[e.g.][]{babu2007}, visualization 
techniques~\citep{comparato2007,way2011,hassan2011} and advanced signal 
processing~\citep{scargle2003,protopapas2006}. All these fields are currently the subject of a new 
discipline: the Astroinformatics~\citep{borne2011}.

In this paper, we have focused our attention on the broad question of how efficiently 
the physical nature of 
astronomical sources can be characterized by multi-wavelength photometric data. We have applied 
CLaSPS to two datasets representing specific cases where such assumption can be tested and verified. 
CLaSPS assumes that low dimensional patterns in data are associated with aggregations (clusters) 
in the structure of the data in the high-dimensional ``\emph{feature} space" generated by all the 
observables of the source\footnote{In general, any source with N measured observables can 
be represented as a point in a N-dimensional \emph{feature} space, where the coordinates
are the numerical values of the observables (or derived quantities).}. 
These clusters are defined by the degree of correlation between the distribution 
of \emph{features} (i.e., the observables used to build the \emph{feature} space where clusters have 
been selected) and a set of external quantities, usually observables, metadata or \emph{a priori} 
constraints that have not been used for clustering. 

The CLaSPS method, based on the KD techniques for unsupervised clustering and the use of external 
information to label the clusters members, has been designed to tackle the problem of the extraction of 
information from two distinct classes of datasets:
a) inhomogeneous large area datasets. The advancements in the Virtual Observatory (VO) technology 
are facilitating the access to datasets obtained by the combination of multiple observations from 
different surveys with different
observational features (e.g., depth, spatial coverage and resolution, spectral resolution). Such datasets 
are, by construction, inherently incomplete and are affected by the inconsistency of the observational 
features of each set of observations used to create them. We expect these datasets to grow in complexity as 
new data becomes available. KD techniques can facilitate 
the extraction of the available knowledge contained in these ``federated" inhomogeneous samples.
b) Large homogeneous datasets from multi-wavelength surveys of well-defined areas of the sky 
observed with similar depths at different wavelengths. These surveys typically 
yield large samples of sources, complete to a given flux. These datasets span limited but 
well characterized regions of the N-dimensional observable \emph{feature} space. The exploration of 
the structure of the multi-dimensional distribution of sources in the \emph{feature} space may lead to 
the discovery of high dimensional correlations and patterns in the data that have been overlooked 
(or, simply, could not be established) in lower dimensional studies.

This paper is organized as follows: in Sec.~\ref{sec:method} we describe the CLaSPS method, 
in Sec.~\ref{sec:experiment1} its application to the CSC+ dataset, and
in Sec.~\ref{sec:experiment2} its application to a sample of blazars with multi-wavelength photometry 
available. We discuss the future developments of CLaSPS in Sec.~\ref{sec:results}.

\section{CLaSPS}
\label{sec:method}

CLaSPS is based on well established data mining techniques for 
unsupervised clustering. These techniques search for spontaneous and 
inherent aggregations of data in the \emph{feature} space generated by their
observables. Table~\ref{tab:definitions} 
summarizes the terms that will be used below. 
These techniques have been complemented by the use of external data (\emph{labels}). \emph{Labels}
are observables not used for the clustering which can be used to characterize the content of the set of 
clusters or of single clusters. \cite{dabrusco2009} used these techniques for 
the selection of optical candidate quasars from photometric datasets.
They employ as \emph{label} the spectroscopic classification available for a subset of the
photometric sources. This method can be extended to use multiple \emph{labels}, both numerical 
(e.g., fluxes, magnitudes, colors) and categorial (spectral classification flags, morphological types). 
From a methodological standpoint, the two tools required for this KD methodology are:

\begin{enumerate}
	\item[(a)] one or more unsupervised clustering algorithms, to determine multiple sets of partitions of 
	the data. (the specific methods used in this paper are discussed in Sec.~\ref{subsec:clustering});
	\item[(b)] a quantitative measure of the degree of correlation between cluster populations and 
	the values of the \emph{label} associated with the members of the clusters (see Sec.~\ref{subsec:scores} 
	for more details);
\end{enumerate}

\noindent Once multiple clusterings\footnote{The term ``clustering" will be used in this paper to indicate 
one collection of clusters determined on any sample by any clustering algorithm. Multiple clusterings 
determined on the same sample can differ for several properties, namely the number of clusters, the number 
of members of the clusters, etc.} of a dataset in a given \emph{feature} space have been produced, the choice 
of the most interesting partition of the dataset is performed considering a quantitative evaluation of the degree 
of correlation between the distribution of the \emph{label} and the cluster population in each clustering. 
Unlike most classical criteria selection that rely only on the intrinsic statistical properties of the clusterings, 
our method selects clusterings based on both the distributions of \emph{features} and of the associated 
\emph{labels}. The degree of correlation between \emph{features} and \emph{labels} can be 
generically expressed by a numerical quantity (the ``score", see Sec.~\ref{subsec:scores}) 
that can be defined and calculated for every single clustering and cluster. 

\begin{table}
\begin{center}
\caption{Definitions of the KD-related terms used in the paper\label{tab:terms}.}
\begin{tabular}{ll}
\tableline\tableline
Term & Definition 														\\
\tableline
Observation			&	An astronomical source as defined by   				\\
					& 	a vector of its observables 						\\
\emph{Feature}			&	Any observable quantity of a given source	 		\\ 
					&	used to determine a set of clusters					\\
\emph{Feature} space	& 	An abstract space where each sample is				\\ 
					& 	represented as points in a N-dimensional space		\\		
Cluster				& 	A set of sources (or observations) aggregated			\\
					& 	by a generic clustering algorithm					\\
Clustering				&	A set of clusters representing a complete 				\\
					&	partition of a sample of observations					\\
Unsupervised			&	In KD, unsupervised clustering refers to the 			\\
clustering				& 	techniques used to determine the spontaneous 		\\
					& 	aggregations of sources not using examples			\\			
\emph{Label}			&	An observable of a set of observations used to			\\
					&	label the members of the clusters (as defined 			\\
					&	in this paper)									\\
Score				&	A quantitative diagnostic of the correlation 			\\
					&	between cluster membership and the distribution  		\\
					&	of \emph{labels} (as defined in this paper)			\\
\tableline
\label{tab:definitions}
\end{tabular}
\end{center}
\end{table}

\subsection{CLaSPS and cluster ensembles}
\label{subsec:comparison}

The task of combining multiple clusterings into a single 
partition is known, in the statistical/data mining literature, 
as the search for the ``consensus clustering". This problem has been thoroughly studied, and 
is discussed in several papers (e.g.~\citet{ghosh2011}). The main reasons for the use of cluster
ensembles techniques are: the improvement of the quality of the clustering, increased robustness of the 
clustering and the ability to combine ``multiview'' clusterings (i.e. of clusterings obtained with nonidentical
sets of sources and/or \emph{features}) \citep{ghosh2011}. 
CLaSPS selects the optimal clustering(s) from the point of view of the astrophysical interpretation of the 
correlations, according to the values of the \emph{scores} (Sec.~\ref{subsec:scores}). The \emph{scores} 
are evaluated on the basis the clustering memberships and a given partition the \emph{label}, 
an external quantity not 
used to produce the clusterings. For this reason, CLaSPS neither tries to determine a 
``consensus clustering" nor attempts to combine clusterings or improve the properties of each distinct  
clustering produced by the UC methods used. 
All the clusterings retain their own properties, biases and weaknesses, that have to be taken into
account when interpreting the results of the application of CLaSPS. The CLaSPS method could be 
nonetheless improved by the application of cluster ensembles techniques, as discussed in Sec.~\ref{sec:results}.

\subsection{Unsupervised clustering}          
\label{subsec:clustering}

In statistical terms, the cluster analysis of a sample is the determination of a segmentation of the data 
in groups 
or clusters, each group representing objects with similar properties \citep{hastie2009,hartigan1975}. The 
cluster analysis depends on the definition of a ``dissimilarity" employed to assign the objects to different 
clusters. Usually, the dissimilarity is evaluated on general attributes (or \emph{features}) 
of the objects. E.g., the values of the observed fluxes in a given filter represent one of the \emph{features} 
of an astronomical dataset. The pairwise dissimilarity between 
the $i$-th and $k$-th observations on the values of the $j$-attribute can be defined as:

\begin{equation}
D(x_i, x_k) = \sum_{j=i}^{p} d_j(x_{ij}, x_{kj})
\end{equation} 
 
\noindent For quantitative attributes, the pairwise dissimilarity can be evaluated using the squared distance 
$d_j(x_{ij}, x_{ik})\!=\!(x_{ij}\!-\!x_{ik})^2$. The individual dissimilarities evaluated for each attribute are 
then combined 
to produce a single overall dissimilarity between objects. The goal of the clustering algorithm is 
to partition the sources into clusters so that the pairwise dissimilarities between objects assigned to 
the same cluster are generally smaller that those in different clusters. In KD the term unsupervised 
refers to learning algorithm that do not require an example or a ``teacher" to infer the properties of the 
probability density associated with a given dataset~\citep{hastie2009}. 
The use of unsupervised techniques is relatively new in astronomy, while supervised learning techniques 
are very common and are usually applied to classification and regression 
problems. An early example of an application of unsupervised clustering to the problem of 
the classification of gamma-ray bursts can be found in~\citep{mukherjee1998}. The
estimation of photometric redshifts for extragalactic sources, based on the spectroscopic redshifts 
measured for a subset of the sources has been tackled with several distinct KD methods, for example
connectivity analysis~\citep{freeman2009}, gaussian processes~\citep{bonfield2010,way2006} and 
neural networks~\citep{yeche2010,collister2004}.
A further example of the combined use of unsupervised clustering and supervised learning techniques 
for photometric redshifts estimation can be found in~\citep{laurino2011}. In general, unsupervised learning 
can be used to highlight the intrinsic 
structure of the data and as an exploratory tool. 
For some of these techniques, the only information
that has to be provided before the clustering is performed is the final number of clusters. In the following 
subsections, we will shortly describe the three unsupervised clustering algorithms used in this work.

\subsubsection{K-means}
\label{subsubsec:kmeans}

The K-means algorithm \citep{lloyd1957} is one of the most frequently used clustering methods. It is 
applicable when the attributes are quantitative and the dissimilarity measure is defined 
as the squared euclidean
distance: $d(x_i, x_k)\!=\!\sum_{j=1}^{p}(x_{ij}\!-\!x_{kj})^2\!=\!||x_{ij}\!-\!x_{kj}||^2$. The N observations 
are associated with K clusters so that in a cluster the average dissimilarity of the observations from the 
cluster mean is 
minimized. $C^*$ is the optimal clustering and $N_k$ is the number of observations assigned to the
$k$-th cluster, defined as:

\begin{equation}
C^{*}\!=\!\mathrm{min}\!\sum_{k=1}^K N_{K}\!\sum_{C(i)=k}\!||x_i\!-\!x_k||^2
\end{equation}

\subsubsection{Hierarchical clustering}
\label{subsubsec:hierarchical}

The result of the application of the K-means clustering technique (Sec.~\ref{subsubsec:kmeans}) to a 
dataset depends on $K$, the number of clusters to be searched. Hierarchical clustering
methods do not require this number to be specified, instead they do require the user to specify a measure of
dissimilarity between groups of observations based on the pairwise dissimilarities among the 
observations in the two groups. Overall, there are two strategies for hierarchical clustering: agglomerative 
and divisive.
 
In the agglomerative strategy, the algorithm starts aggregating at the lowest level possible 
(each group is composed by only one observation) and at each level (or generation), a pair 
of clusters is recursively merged into a single cluster. In the divisive strategy, the starting point is 
the top of the ``tree" (all 
observations in one cluster) and at each level each cluster is recursively split into two new clusters. 
The merging in the agglomerative method, at each level, involves the aggregation of the two groups 
with the smallest intergroup dissimilarity. In the divisive methods, instead, at each level the splitting 
produces two new clusters with the largest possible between-group dissimilarity. 
Recursive splitting/agglomeration can be represented by a rooted binary 
tree, where the nodes represent groups. The root node is associated with the whole dataset and each 
terminal node represents one of the individual observations. A common representation of the 
hierarchical structure 
called dendrogram~\citep{hartigan1975} is obtained by plotting the binary tree so that the height of each 
node is proportional to the value of the intergroup dissimilarity between its ``children" nodes. Hierarchical 
clustering techniques impose a hierarchical structure to the 
data even when such structure does not exists in the data. For this reason, in this paper we will not 
interpret the clusterings in terms of the properties of the hierarchical structure they belong to, but 
only in terms of the properties of their \emph{feature} distribution and of the properties of the distribution of 
the \emph{labels} associated with the cluster members.

\noindent Agglomerative hierarchical clustering depends on: a) the choice of the definition of pairwise 
dissimilarity (i.e., dissimilarity between the members of a pair of observations), and b) the agglomeration 
or ``linking" strategy, i.e. the definition of the inter-group dissimilarity, usually based on the pairwise 
dissimilarity adopted. Several pairwise dissimilarity definitions have been used for the method 
described in this paper; these include the Euclidean distance 
(see Eq.~\eqref{eq:euclidean}), the Manhattan distance (Eq.~\eqref{eq:manhattan}) (also known as the 
taxicab distance) and the maximum (or Chebyshev's) distance 
(Eq.~\eqref{eq:maximum})\footnote{These three 
distances are special cases of the general Minkowski's distance defined as:
	\begin{equation}
		D(x_{i}, x_{k}) = \sqrt[p]{\sum_{j=1}^{N} (x_{ij}\!-\!x_{kj})^{p}}\label{eq:minkowski},
	\end{equation}
for values of the parameter p equal to 2, 1 and $\infty$ respectively.}

\begin{eqnarray}
D(x_{i}, x_{k}) = \sqrt{\sum_{j=1}^{N} (x_{ij}\!-\!x_{kj})^{2}}\label{eq:euclidean}\\
D(x_{i}, x_{k}) = \sum_{j=1}^{N} (\|x_{ij}\!-\!x_{kj}\|)\label{eq:manhattan}\\
D(x_{i}, x_{k}) = \mathrm{max}_{i}(\|x_{ij}\!-\!x_{kj}\|)\label{eq:maximum}
\end{eqnarray}

\noindent All the above metrics are suited for continuous measurements associated with the observations. 
The linking strategies used in this work are described below. The descriptions of the distinct linkage
strategies are given in the case of agglomerative clusterings, but they are also valid for divisive 
clustering methods:

\begin{description}
\item[a)] Single linkage: the inter-clusters dissimilarity between two generic clusters $A$ and $B$ can be 
defined as the minimum pairwise dissimilarity between observations of each cluster:
	\begin{equation}
	D(A, B)\! = \! \mathrm{min}\{D(x, y): x\in\!A, y\in\!B\}
	\end{equation}
The clusters are grouped on the basis of the closest couple of members. For this reason, clusters which, 
on average, are not the closest but which share few nearby observations, can be merged. This is similar to 
what happens in clustering methods based on the ``friends-of-friends" algorithm~\citep{hartigan1975};
\item[b)] Complete linkage: the inter-cluster dissimilarity between two clusters $A$ and $B$ can be 
defined as the maximum pairwise dissimilarity between observations belonging to the two clusters, namely:
	\begin{equation}
	D(A, B)\! = \! \mathrm{max}\{D(x, y): x\in\!A, y\in\!B\}
	\end{equation} 
In this case, clusters are merged when globally very close to each other, since the condition is on the farthest
pair of observations. 
\item[c)] Average linkage: the inter-clusters dissimilarity between the two clusters $A$ and $B$ can be 
defined as the pairwise dissimilarity between the average observations for each clusters:
	\begin{equation}
	D(A, B)\! = <\!(D(x, y): x\in\!A, y\in\!B)\!>
	\end{equation}
This case is intermediate between the single linkage and complete linkage strategies. Clusters are 
merged when 
they are on average close to each other, i.e. most of observations of each cluster are close to each other. This
strategy produces the stablest configuration because it is not sensitive to ``outliers".	
\item[d)] Ward's linkage: the inter-clusters dissimilarity can be defined as a measure of the 
increase of variance of the cluster obtained by merging the parents clusters compared to the sum 
of the variances of the two separate clusters:
	\begin{equation}
	D(A, B)\! =\! \mathrm{ESS}(A, B) - [\mathrm{ESS}(A) \!+\! \mathrm{ESS}(B)]
	\end{equation}
where:
	\begin{equation}
	\mathrm{ESS}(A)\! =\! \sum_{i=1}^{N_{A}}\|x_{i} -\frac{1}{N_{A}}\sum_{j=1}^{N_{A}}x_{j}\|^{2}
	\end{equation}
This linkage strategy provides compact and spherical clusters which, intrinsically, have minimal internal 
variance~\citep{ward1963}. 	
\end{description}


\subsubsection{Self Organizing Maps}
\label{subsubsec:som}

Self Organizing Maps (SOM)~\citep{kohonen1990,vesanto2000} are a constrained 
version of the K-means clustering. In SOM, the ``prototypes", template observations 
determined on the basis of the initialization of the algorithm, are encouraged to lie 
on a two-dimensional surface. This manifold is called
a \emph{constrained topological map}, since all the  observations in the original \emph{feature} space
are mapped to a two-dimensional coordinate system. A two dimensional grid of prototypes is ``bended"
by the SOM algorithm to adapt to the observations as accurately as possible. Once reached optimal
mapping, the observations can be mapped down onto the ``prototypes", and each observation is assigned 
to the cluster
represented by the closest prototype. Given the $K$ prototypes $m_k$ and the $i$-th observation in the
p-dimensional \emph{feature} space $x_i$, the closest prototype $m_j$ is picked using euclidean distance 
\eqref{eq:euclidean}. In the simplest version of the SOM, the position of the prototype is updated according 
to the rule:

\begin{equation}
m_k \leftarrow m_k + \alpha (x_i - m_k)
\label{eq:som}
\end{equation}

\noindent where $\alpha$ is a number called \emph{learning rate} that changes at each iteration and, 
usually, goes from $\sim 1$ to 0 over few thousands iterations. The positions of the prototypes are updated
until the distance of each observation associated with prototype becomes smaller that a given 
``distance threshold" 
$r$. The distance threshold $r$ decreases linearly with each new observation considered, according to  
the empirical rule followed to update the value of $r$. As in the case of the K-means clustering, 
the number of prototypes, i.e. of final number of clusters, must be specified by the user. SOM becomes an 
online\footnote{An online algorithm is one that can process its input in a piece-by-piece fashion, so that the
whole input is not available from the start.} version of the K-means algorithm for small $r$ distances, yielding 
only one observation 
associated with each prototype. The SOM algorithm can also be used 
as a supervised classification and regression method, using the stable prototypes definitions obtained 
using a \emph{training}, so that ``new" sources will be associated with the closest prototype in the 
\emph{feature} space. Because of their versatility, SOM have recently 
been applied to astronomical data to address distinct problems: the selection and classification of 
extragalactic sources from large surveys data using 
their photometric attributes~\citep{geach2012}, the evaluation of photometric 
redshifts~\citep{geach2012,way2012}, spectral classification of stars~\citep{bazarghan2012} and the
reconstruction of large scale structure of galaxy distribution~\citep{way2011}.

\subsection{The Score}
\label{subsec:scores}

As discussed in Sec.~\ref{sec:method}, the originality of the CLaSPS method lies in the criterion 
used to select the most meaningful aggregations of sources in the \emph{feature} space, which 
is based on the correlation with the \emph{labels}, i.e. 
other observables not used to produce the clustering. This correlation 
is evaluated using a novel indicator, called the \emph{score}. 

Each one of the methods described in the previous paragraphs 
provides us with one or more clusterings when applied to a given dataset. Each observation is 
uniquely associated with one of the clusters in the clustering (i.e., each observation belongs to one and 
only one cluster for each clustering). Additional information available for each source
in the dataset, but not used for the clustering (i.e. not used to build the \emph{feature} space) can be used to 
label the content of the clusters of each clustering. Categorial \emph{labels}
provide a natural binning; continuous \emph{labels}, in our method, must be binned for the evaluation 
of the \emph{scores}. The binning can be either a set of continuous intervals 
(for continuous \emph{labels}) or a set of (single or grouped) values (for 
categorial \emph{labels}). The distribution of \emph{labels} values for the members of each cluster 
is used to determine the level of correlation between the \emph{label} and the single cluster. The 
degrees of correlation between the \emph{label} distribution and each cluster of a clustering are then
combined to provide a measure of the degree of correlation of the \emph{label} distribution with the
clustering as a whole. 

The \emph{score} provides a quantitive measure of the 
correlation between a \emph{label} and the cluster membership for a given clustering. 
For any \emph{label} $L$ in the set of $N_{L}$ \emph{labels} 
available, a binning of $L$ is represented by a set of $M^{(L)}$ classes, either quantitative intervals or 
categorial values $\{C^{(L)}_{1}, C^{(L)}_{2},..., C^{(L)}_{M}\}$. 
The basic element of the \emph{score} definition is the fraction $f_{ij}$ of the $i$-th cluster 
members with values of the \emph{label} falling in the $j$-th class:

\begin{equation}
f_{ij}=\frac{n_{ij}(L_{i} \in C^{(L)}_{j})}{N_{i}}
\label{eq:fraction}
\end{equation}

\noindent where $L_{i}$ is the set of values of the \emph{label} $L$ associated with the members of 
the $i$-th cluster, $n_{ij}$ is the number of members of the $i$-th cluster with \emph{label} values 
belonging to the $j$-th \emph{label} class, and $N_{i}$ is the number of members of the $i$-th 
cluster of the clustering. $F_{ij} = \{F_{i1}, F_{i2},...,F_{iM^{(L)}}\}$ can be defined as the arrangement 
of the fractions $f_{ij}$ sorted in increasing order\footnote{A distinct arrangement of the fractions 
$f_{ij}$ is determined for each cluster.}. The \emph{score} of the $i$-th cluster of a clustering 
composed of $N_{\mathrm{clust}}$ clusters relative to the \emph{label} $L$ can be defined as 
follows:

\begin{equation}
S_{i}\!=\!\sum_{j=2}^{M^{(L)}}\|F_{ij}\!-\!F_{i(j\!-\!1)} \|
\label{eq:scoreclust}
\end{equation}

\noindent Using the definition of \emph{score} for a single cluster, the total \emph{score} of the 
clustering relative to the \emph{label} $L$ can be defined as:

\begin{equation}
S_{\mathrm{tot}} \! = \! \frac{1}{N_{\mathrm{clust}}}\!\times\!\sum_{i=1}^{N_{\mathrm{clust}}} S_{i}\!=
\!\frac{1}{N_{\mathrm{clust}}} 
\sum_{i=1}^{N_{\mathrm{clust}}} 
\!\big(\!\sum_{j=2}^{M^{(j)}} \! \| F_{ij} - F_{i(j \! - \! 1)} \! \| \big)
\label{eq:scoretot}
\end{equation}

\noindent where $S_{i}$ is the \emph{score} evaluated for the $i$-th cluster of the clustering defined 
in \eqref{eq:scoreclust}. By definition, the total \emph{score} $S_{\mathrm{tot}}$ and each single 
cluster \emph{score} $S_{i}$ are normalized to unity. 

The weighted total score $S_{\mathrm{tot}}^{'}$ can be defined as the total normal score $S_{\mathrm{tot}}$
where the \emph{score} of each cluster is weighted according to the number of sources of the 
cluster:

\begin{equation}
S_{\mathrm{tot}}^{'}\!=\!\frac{1}{N_{\mathrm{clust}}}\frac{\sum_{i=1}^{N_{\mathrm{clust}}}N_{i}\!\times\!S_{i}}{\sum_{i=1}^{N_{\mathrm{clust}}}N_{i}}
\!=\!\frac{1}{N_{\mathrm{clust}}}\frac{\sum_{i=1}^{N_{\mathrm{clust}}}N_{i}\!\times\!S_{i}}{N_{\mathrm{tot}}}
\label{eq:weightedscoretot}
\end{equation}

\noindent where $N_{\mathrm{tot}}$ is the total number of sources in the clustering. 

The contributions of all clusters to the total $S_{\mathrm{tot}}$ \emph{score} are equally weighed. 
For this reason, $S_{\mathrm{tot}}$ is sensitive to 
small clusters with few members with a large degree of correlation with the \emph{label} (for example 
singletons, i.e. clusters composed by only one observation). Instead, the contribution of each 
cluster to the weighted total \emph{score} $S_{\mathrm{tot}}^{'}$ is proportional to the ratio of its members 
to the total number of observations in the clustering. By definition, $S_{\mathrm{tot}}^{'}$ 
is a declining function of the total number of clusters of the clustering.
The weighted total \emph{score} $S_{\mathrm{tot}}^{'}$ is heavily influenced by 
the largest clusters and, as a consequence, is a measure of the ``mass-weighted" degree of 
correlation of the dataset. Both $S_{\mathrm{tot}}$ and $S_{\mathrm{tot}}^{'}$ are used to select 
the optimal clusterings because they represent complementary measures of distinct 
aspects of the clusterings, namely the level of correlation of the largest clusters and of the 
existence of less populated groups of sources.

\subsubsection{Score assessment}
\label{subsubsec:simulations}

Before applying the \emph{scores} to the real astronomical datasets described in 
Sec.~\ref{sec:experiment1} 
and Sec.~\ref{sec:experiment2}, we have tested the effectiveness of this method with simulated 
clusterings. In these simulations, we assume that the final structure of a generic clustering is independent 
from the unsupervised clustering algorithm used to produce the clustering. This assumption is 
reasonable because the algorithm depends 
on the topological relations among the sources of the dataset in the \emph{feature} space where the 
clustering has been generated and on the linking strategy used to associate the sources (see 
Sec.~\ref{subsec:clustering} for more details). Both the properties of the clustering algorithm and the 
actual values of the \emph{features} and \emph{labels} associated with the simulated sources are 
of no importance in an idealized description of the clustering, where topological and 
relational properties of the observations are condensed in the membership, a categorial information, for
each source. On these premises, the fundamental parameters describing simulated clusterings 
are: the total number 
$N^{\mathrm{(sim)}}_{\mathrm{Tot}}$ of observations of the sample of the simulated clustering; the number of 
clusters in the clustering $N_{\mathrm{clust}}^{\mathrm{(sim)}}$; the number of members of each cluster, 
normalized to the total number of observations in the sample $N_{\mathrm{i}}^{\mathrm{(sim)}}$ 
where $i \in \{1,...,N_{\mathrm{clust}}^{\mathrm{(sim)}}\}$ is associated with the spread of the sizes of the 
clusters measured with the variance $\sigma^{2}(N_{i}^{(\mathrm{sim})})$; the number of classes of the 
\emph{label} $M_{L}^{\mathrm{(sim)}}$; and a prescription to assign the \emph{label} values to the 
cluster members. 

Three different scenarios have been considered in order to create realistic simulated clusterings.
These scenarios have inspired distinct association rules between classes of \emph{label} 
values and observations in the clusters that have been used to generate the simulated clusterings. 
These scenarios are: a) \emph{label} values belonging to any \emph{label} class are randomly associated 
with the 
observations, regardless of their membership; b) \emph{label} values of each \emph{label} class are 
assigned only to 
sources belonging to a fixed small number of randomly selected clusters; and c) \emph{label} values of each 
\emph{label} class are assigned to observations in only one randomly selected cluster in the clustering. 

We produced simulated clusterings with different degrees of correlation between \emph{label} 
values and clusters memberships by mixing the previous three prescriptions in 
different percentages. Thus, we obtained ``recipes" to simulate weak, moderate and strong correlations
in clusterings.
For the fraction of class \emph{label} values not randomly associated, each value was 
assigned only to sources belonging to a randomly picked number of clusters, with one, two or three 
clusters being the most likely options by definition. The remaining fraction of \emph{label} values was 
randomly assigned to sources of the clusterings independently from their cluster memberships.  

The weakly correlated clusterings have a 0\%-30\% not randomly assigned class, the moderately 
correlated clusterings have 30\% to 70\% and the extremely correlated clusterings have 70\% to 100\%.
These intervals have been selected in order to verify the effectiveness of the \emph{score} to express the 
level of correlation among \emph{label} distribution and cluster memberships in realistic scenarios where 
the \emph{classes} are partially correlated with a subset of clusters, and 
in the extreme cases with total correlation (100\% of not randomly assigned classes) and no
correlation (0\% of not randomly assigned classes), as a function of the parameters of the simulated 
clusterings.
We produced equal numbers of simulated clusterings for each of the three prescriptions described above. 
All the parameters of the simulations were free to vary in the intervals described in 
Table~\ref{tab:simulations}, where also the composition of the three classes of clusters is summarized.

\begin{table}
\begin{center}
\caption{Ranges of the parameters of the simulated clusterings used to validate the effectiveness
of the \emph{score} definitions in capturing the degree of correlation between \emph{label} value classes 
and clusters membership distributions\label{tab:simulations}}
\begin{tabular}{cc}
\tableline\tableline
Parameter								&  	Value(s)		\\
\tableline
$\mathrm{N}^{\mathrm{sim}}$					&	1000 		\\
$\mathrm{N}^{\mathrm{sim}}_{\mathrm{Tot}}$		&	[30, 200] 		\\
$\mathrm{N}^{\mathrm{sim}}_{\mathrm{clust}}$		&	[3, 12] 		\\
$\mathrm{M}^{\mathrm{sim}}_{\mathrm{(L)}}$		&	[2, 10]	 	\\
\tableline
\\
\\
\\
\tableline\tableline
Type of clustering							&  	\% not-random/random \\
\tableline
Weakly/not corr.							&	[0\%, 30\%],[100\%, 70\%]  \\
Moderately corr.							&	[30\%, 70\%],[70\%, 30\%]  \\		
Strongly corr.								&	[70\%, 100\%],[30\%, 0\%]  \\	
\tableline
\end{tabular}
\end{center}
\end{table}

\begin{figure*}
\begin{center}
\begin{tabular}{cc}
\includegraphics[width=0.5\linewidth]{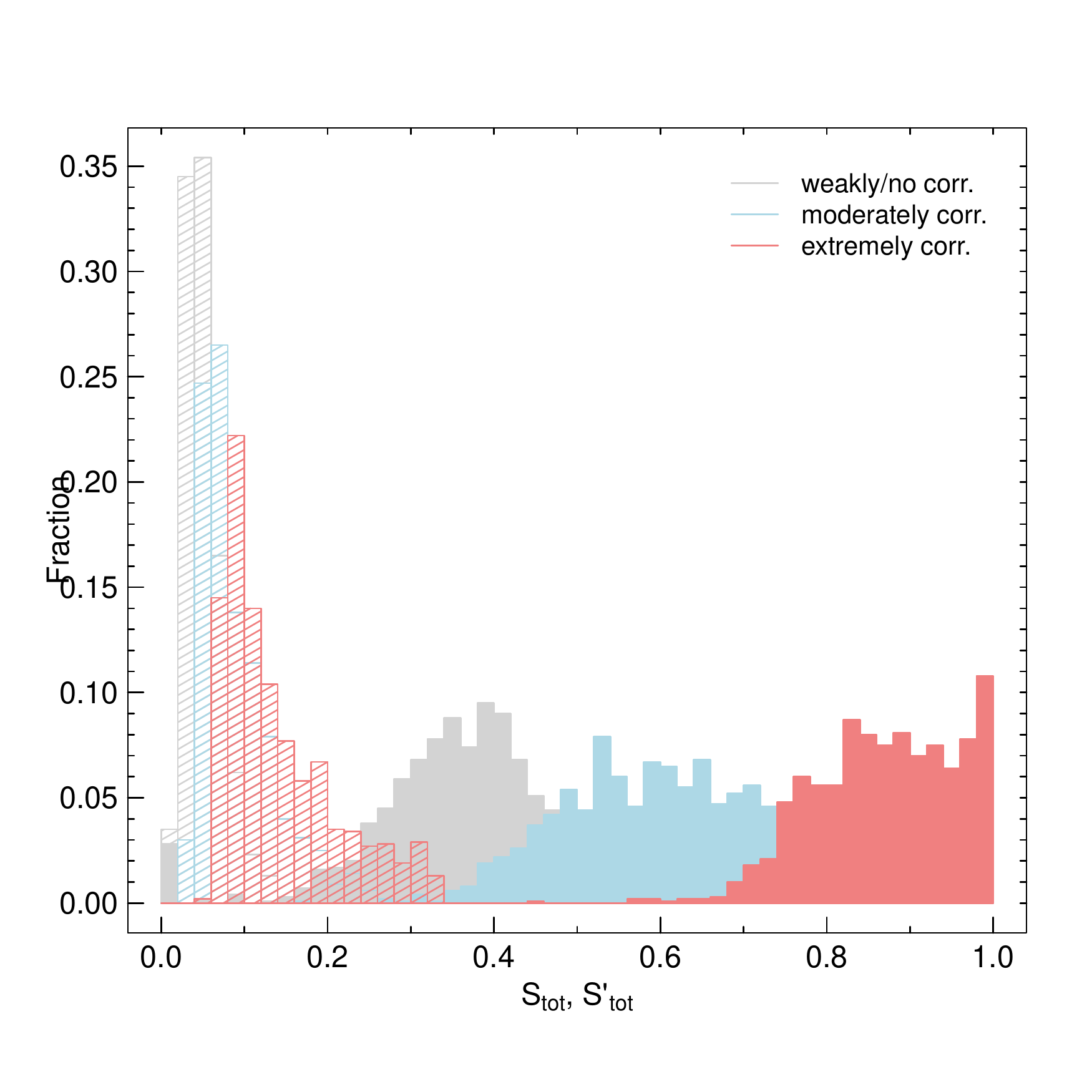}
\includegraphics[width=0.5\linewidth]{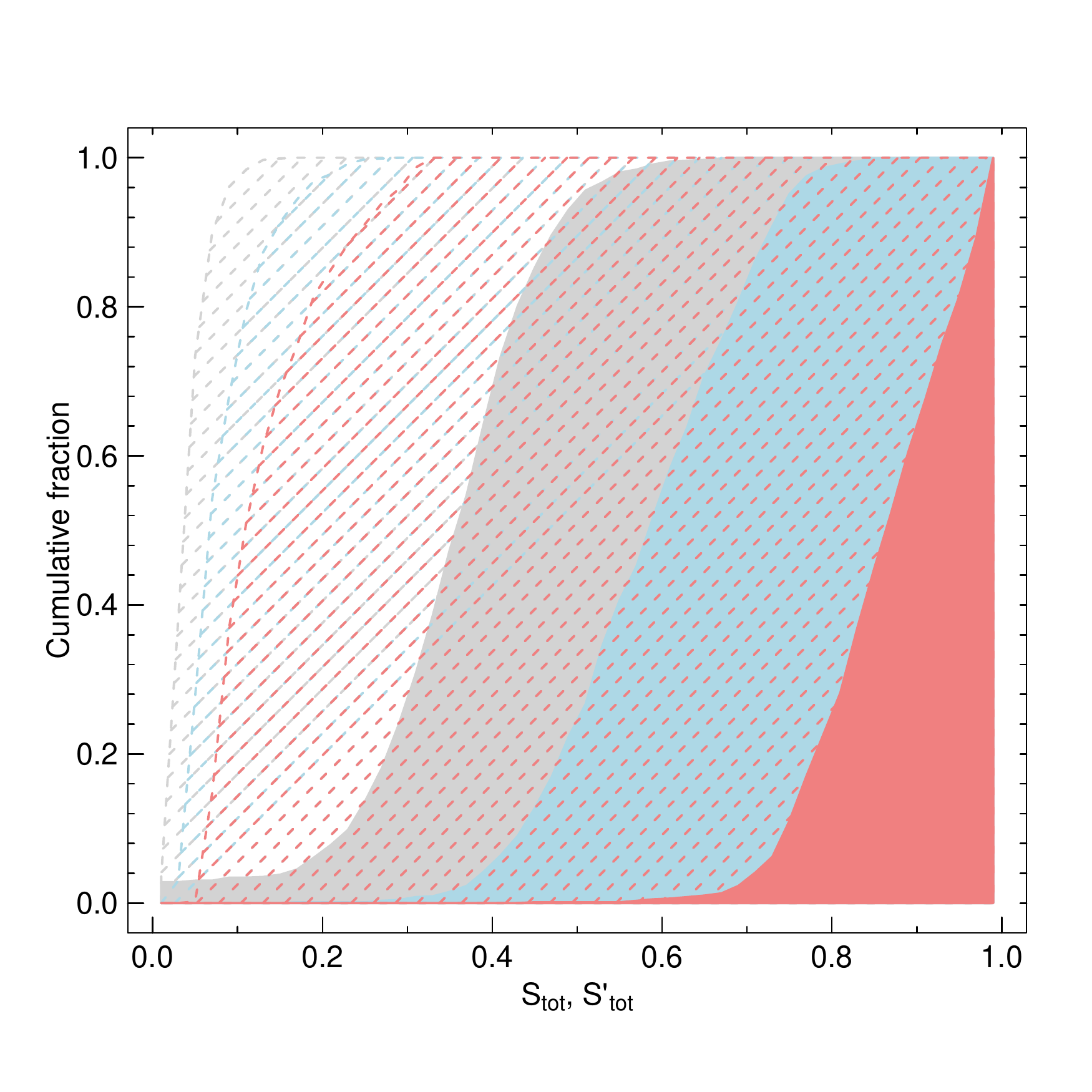}
\end{tabular}
\end{center}
\caption{Left: the normalized histograms of the values of weighted and unweighted \emph{scores} for 
the three classes of simulated clusterings (gray: weakly correlated clusterings, blue: moderately correlated
clusterings, red: extremely correlated clusterings) are shown as solid and open bars 
respectively. Right: cumulative distributions of the \emph{scores} for the three classes of 
clusters are shown.}
\label{fig:histograms}
\end{figure*}

\begin{figure*}
\begin{center}
\begin{tabular}{cc}
\includegraphics[width=0.4\linewidth]{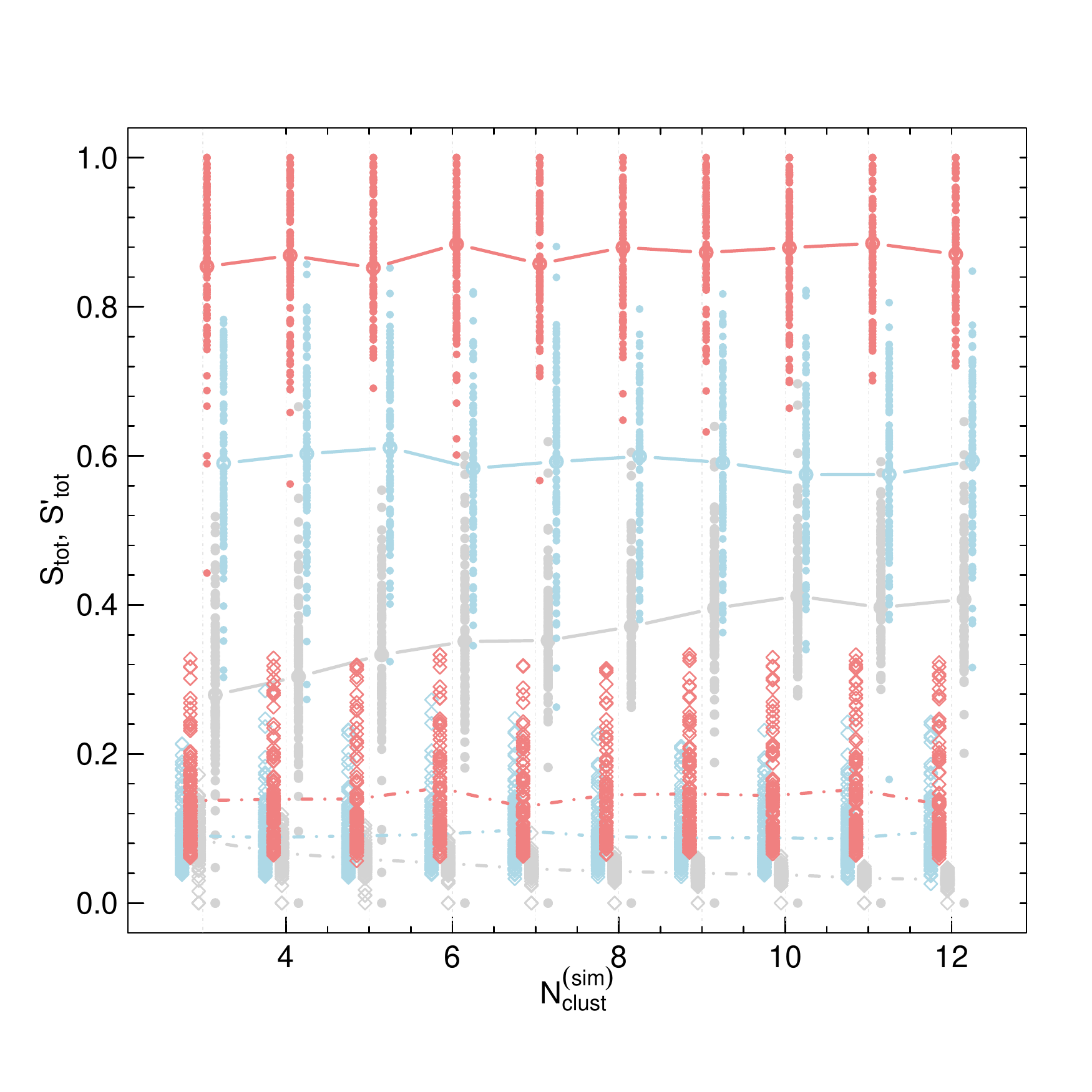} &
\includegraphics[width=0.4\linewidth]{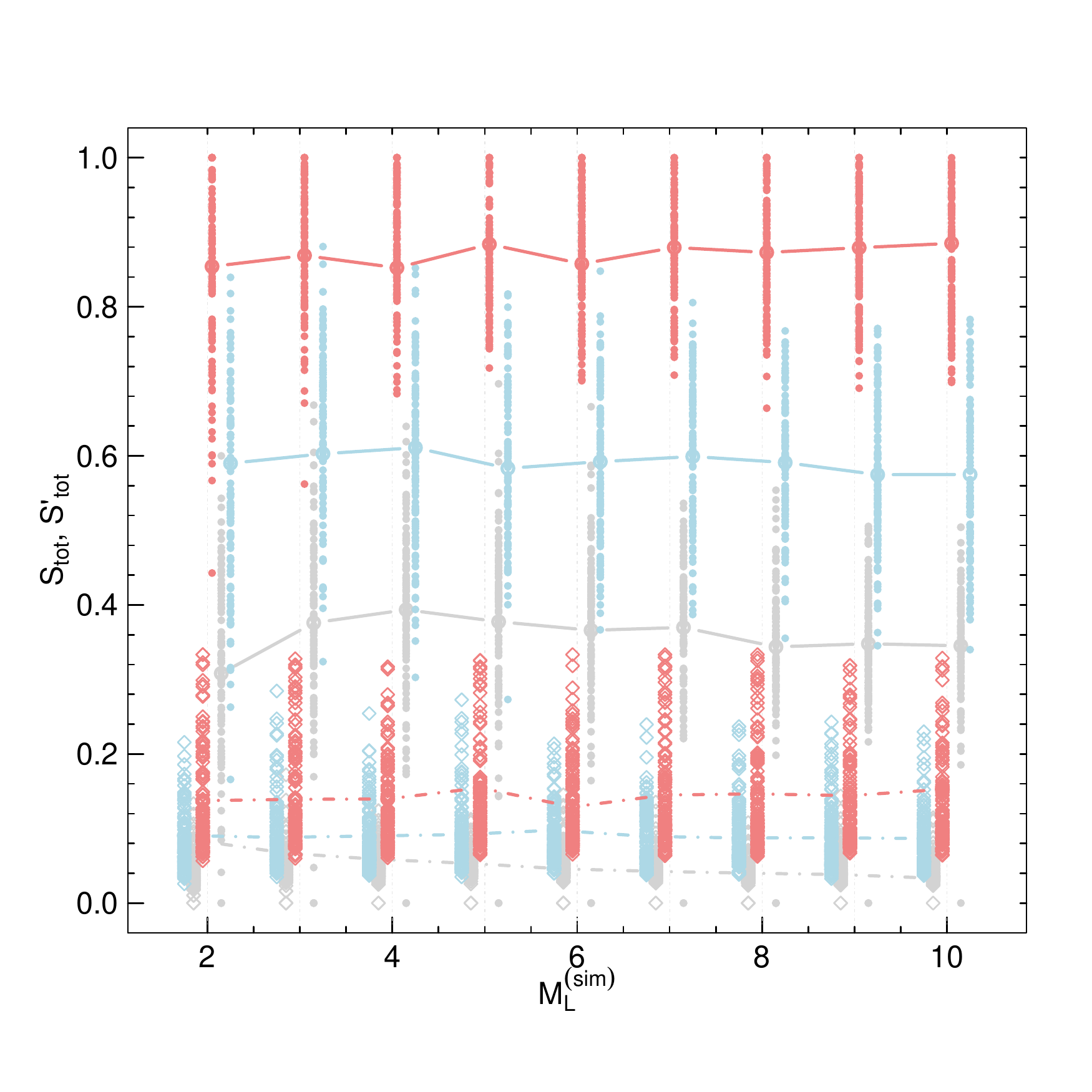} \\
\includegraphics[width=0.4\linewidth]{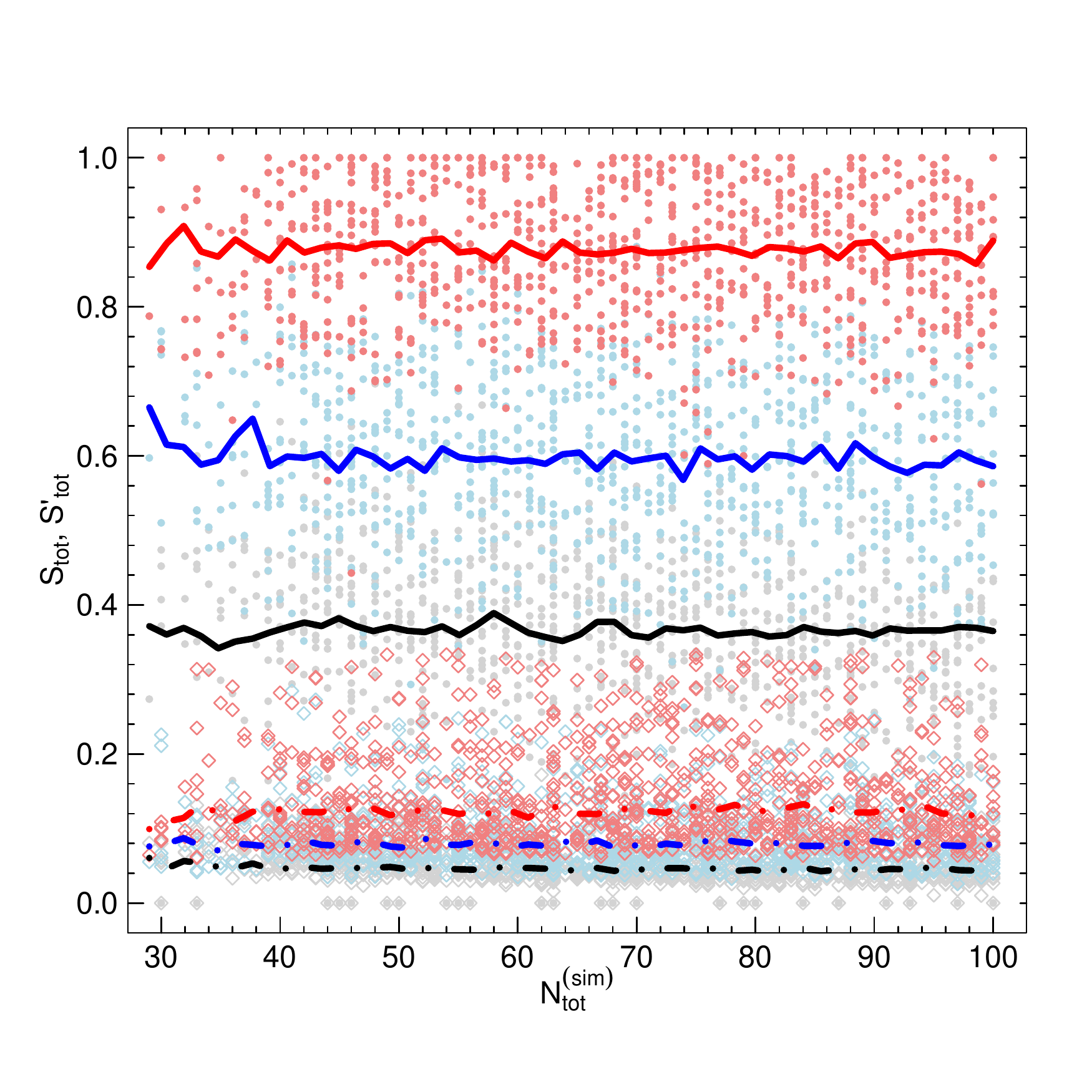} &
\includegraphics[width=0.4\linewidth]{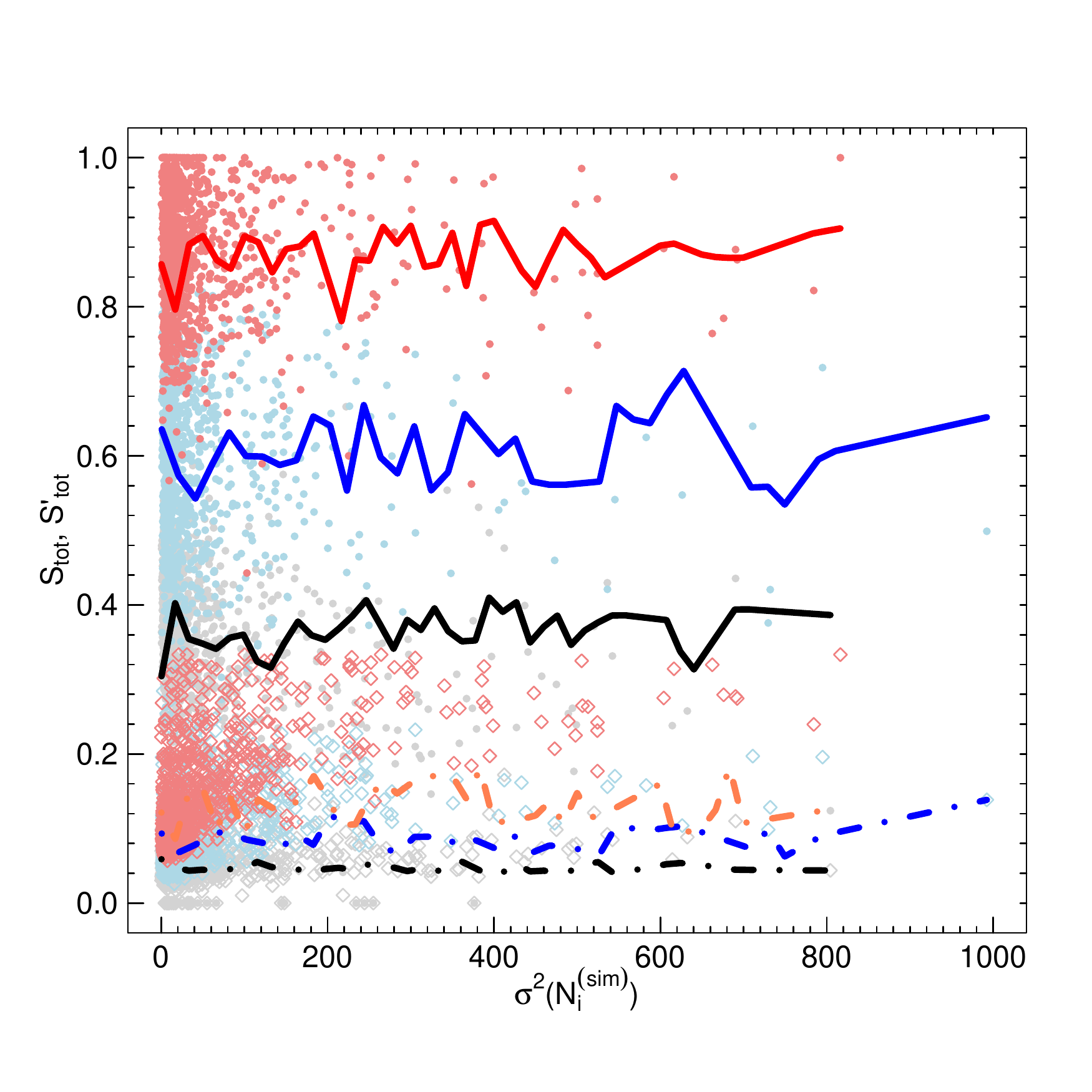}
\end{tabular}
\end{center}
\caption{Plots of the \emph{score} values for simulated clusterings as functions of the general parameters of the 
simulated clusterings. In all plots, the gray, blue and red symbols are associated with 
the \emph{scores} evaluated 
for the three families of clusterings with weakly, moderately and extremely 
correlated association of the values of \emph{label} classes to the members of the clusters. Top left: 
the values of the \emph{scores} are plotted as functions of the number of clusters 
$\mathrm{N}^{\mathrm{(sim)}}_{\mathrm{clust}}$ 
of the simulated  clusterings; top right: \emph{score} values as function of the number 
of classes of the \emph{labels} $M_{L}^{\mathrm{(sim)}}$. Bottom left: \emph{score} values plotted
as function of the total number of observations $N_{\mathrm{tot}}^{\mathrm{(sim)}}$ of the simulated 
clusterings; bottom right: the \emph{score} values as function of the variance of the sizes 
of the clusters belonging to the simulated clusterings. In all plots, weighted and normal \emph{scores} are 
represented by open and
filled symbols respectively, know while the continuous lines show the average 
and to the moving window medians of the \emph{scores} distributions for each class of simulated 
clusterings, in the upper and lower plots respectively.}
\label{fig:simulations}
\end{figure*}

Both normal \emph{scores} (Eq.~\eqref{eq:scoretot}) and weighted \emph{scores} 
(Eq.~\eqref{eq:weightedscoretot}) 
were evaluated in each family of simulated clusterings. The histograms of the distributions of values 
of $S_{\mathrm{tot}}$ and $S_{\mathrm{tot}}^{'}$ for the simulated clusterings are shown in 
Figure~\ref{fig:histograms}. The scatterplots of the values of the \emph{scores} as functions of the parameters of 
the simulations $N_{\mathrm{tot}}^{(\mathrm{sim})}$, $N_{\mathrm{clust}}^{(\mathrm{sim})}$, 
$M_{L}^{(\mathrm{sim})}$ and $\sigma^{2}(N_{i}^{(\mathrm{sim})})$ are shown in Figure~\ref{fig:simulations}.

From the differential and cumulative histograms shown in the plots in 
Figure~\ref{fig:histograms}, it is evident that the normal \emph{score} $S_{\mathrm{tot}}$ (filled bars) spans 
the whole range $[0, 1]$ with the values of the extremely correlated simulated clusterings ranging from 0.6 
to 1, the moderately correlated with \emph{scores} values between $\sim 0.4$ and 0.7, and weakly correlated 
clusterings have values of the total normal \emph{scores} smaller than 0.5. 
In the same plots, the values of the total weighted \emph{scores} 
$S^{'}_{\mathrm{tot}}$ (dashed bars) are consistently smaller than $S_{\mathrm{tot}}$ and are not normalized 
to unity (as remarked in Sec.~\ref{subsec:scores}). The weighted total \emph{scores} $S^{'}_{\mathrm{tot}}$ for 
the three families of simulated clusterings are separated less clearly than in the case of the normal total 
\emph{scores}, as 
the weights depend on the sizes of the clusters $N_{i}^{(\mathrm{sim})}$ and, consequently, the total value of 
the \emph{score} depends on their variance $\sigma^{2}(N_{i}^{(\mathrm{sim})})$ of the sizes of the clusters. 
Even so, the strongly correlated clusterings are associated with values of $S^{'}_{\mathrm{tot}}$ on 
average larger than 
the $S^{'}_{\mathrm{tot}}$ values for partially correlated simulated clusterings and randomly drawn clusterings.
The four plots in Figure~\ref{fig:simulations} show that the values of the \emph{scores} do not show significant 
dependencies over any of the four parameters describing the simulated clusterings: 
$N_{\mathrm{tot}}^{(\mathrm{sim})}$, $N_{\mathrm{clust}}^{(\mathrm{sim})}$, $M_{L}^{(\mathrm{sim})}$ 
and $\sigma^{2}(N_{i}^{(\mathrm{sim})})$. The results of the simulations demonstrate that the \emph{scores} 
as defined in this paper, are unbiased diagnostics of the degree of correlation between the distribution 
of observations in a clustering and the \emph{label} class values.

\subsection{Choice of Clustering}
\label{subsec:choice}

\begin{figure*}
\begin{center}
\begin{tabular}{cc}
\includegraphics[width=0.5\linewidth]{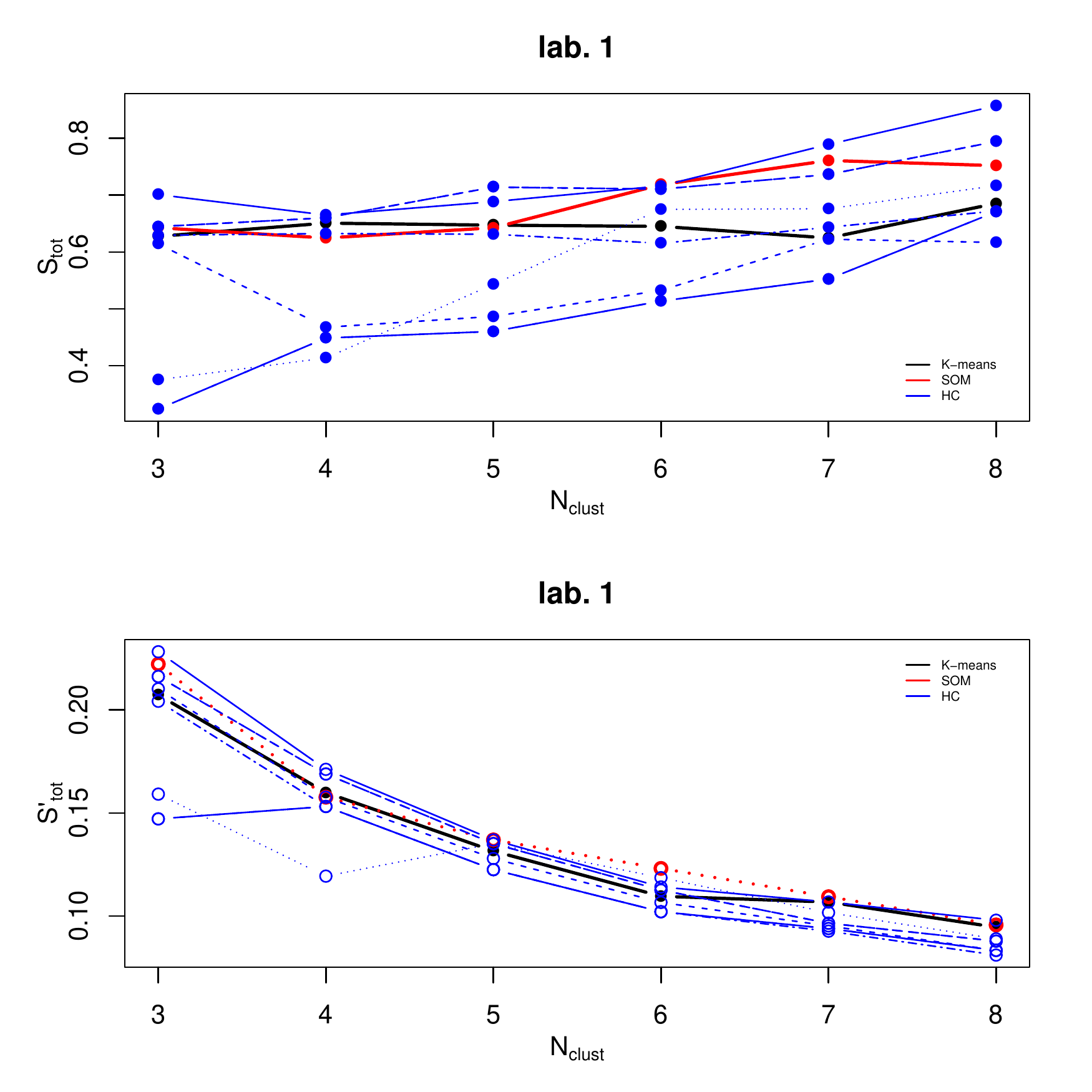}	&
\includegraphics[width=0.5\linewidth]{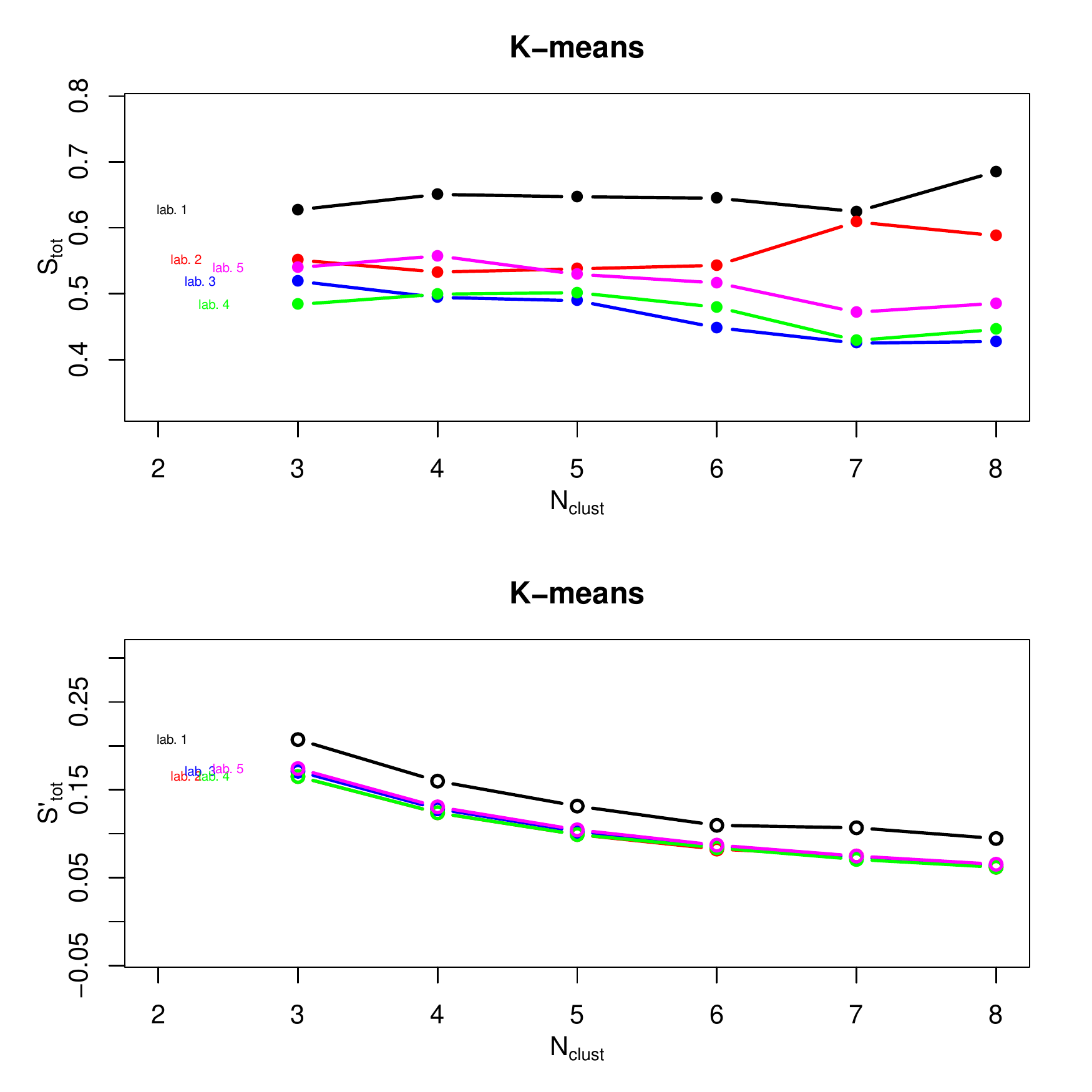}	\\
\end{tabular}
\end{center}
\caption{Plots of the normal and weighted total \emph{score} distributions for a generic dataset for distinct
total number of clusters $N_{\mathrm{clust}}$ of the clustering and type of unsupervised clustering 
algorithm used to produce the clusterings. Left: the total \emph{scores} distributions, evaluated for a specific 
\emph{label} $L$, are plotted as a function of the total number of clusters of the clusterings for all 
unsupervised clustering algorithms; right: both total \emph{score} distributions are plotted 
as a function of the total number of clusters $N_{\mathrm{clust}}$ of the clusterings for multiple \emph{labels}. 
In both plots, the total normal \emph{scores} $S_{\mathrm{tot}}$ and weighted 
\emph{scores} $S^{'}_{\mathrm{tot}}$ are plotted separately for the sake of clarity.}
\label{fig:linear}
\end{figure*}

Given a \emph{label} $L$ and a set of \emph{label} classes $M_{L}$, CLaSPS 
produces distinct values of the total
\emph{scores} $S_{\mathrm{tot}}$ and $S^{'}_{\mathrm{tot}}$ for each clustering produced by 
any clustering method 
employed (see Sec.~\ref{subsec:clustering}). The clusterings produced by a single method differ for the 
total number of clusters $N_{\mathrm{clust}}$. The \emph{scores} can be plotted as a function of the number of 
clusters and the clustering method, as shown in the left side of Figure~\ref{fig:linear}, 
to identify the clusterings with the largest degree of correlation between the \emph{label} classes and the 
clustering members. A similar plot can be used to immediately determine the clustering with the 
largest correlation between the \emph{features} distribution of the clusters and each of the whole set of 
\emph{labels} at once (right side plot in Figure~\ref{fig:linear}).

Since the total \emph{scores} are averaged over all the clusters in a given clustering, 
they can only provide 
information on the global degree of correlation of the distribution of \emph{label} classes and 
the clustering membership. Information about the local correlations is carried by the value of 
the \emph{scores} for each cluster contained in the clustering separately.

\begin{figure*}
\begin{center}
\begin{tabular}{cc}
\includegraphics[width=0.5\linewidth]{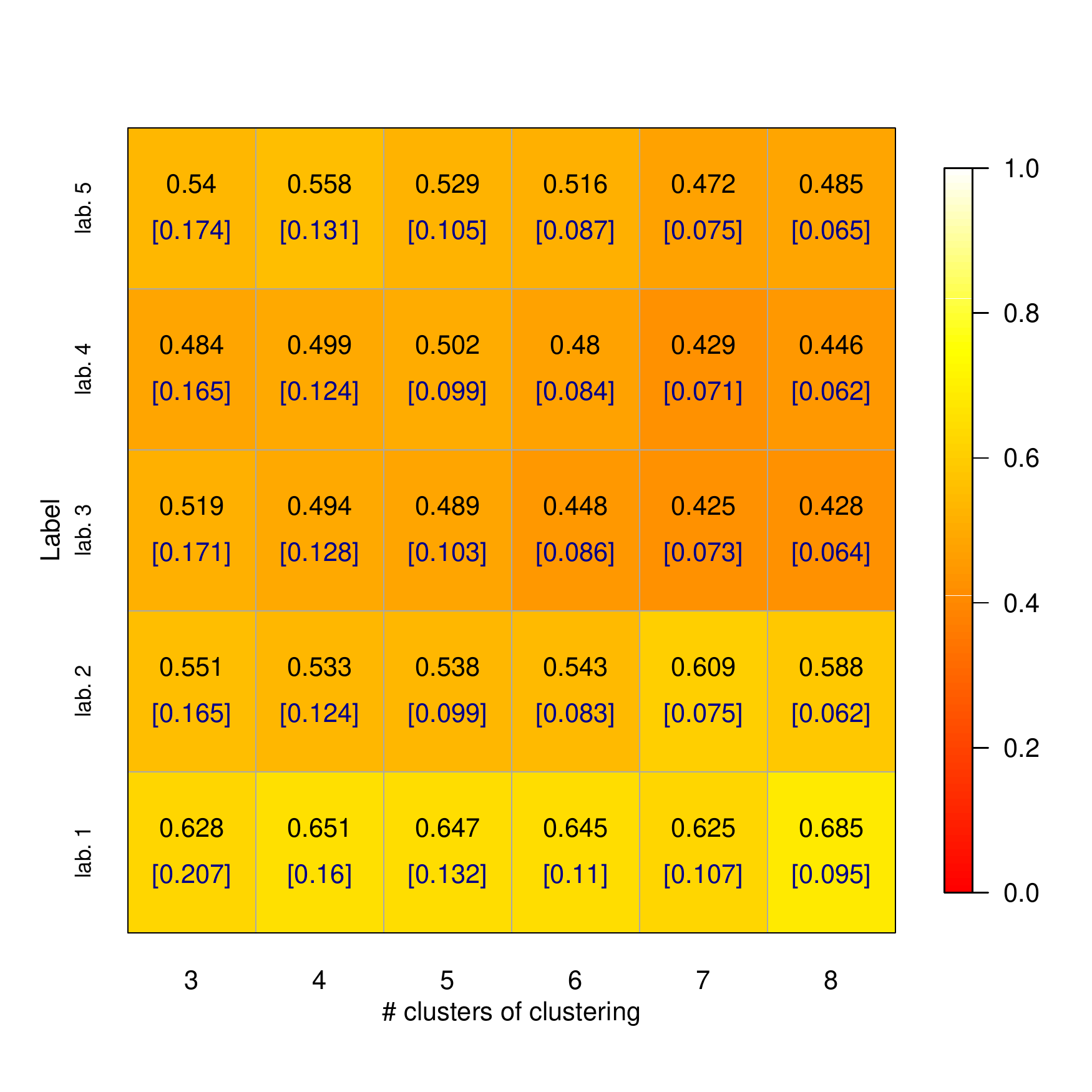}	&
\includegraphics[width=0.5\linewidth]{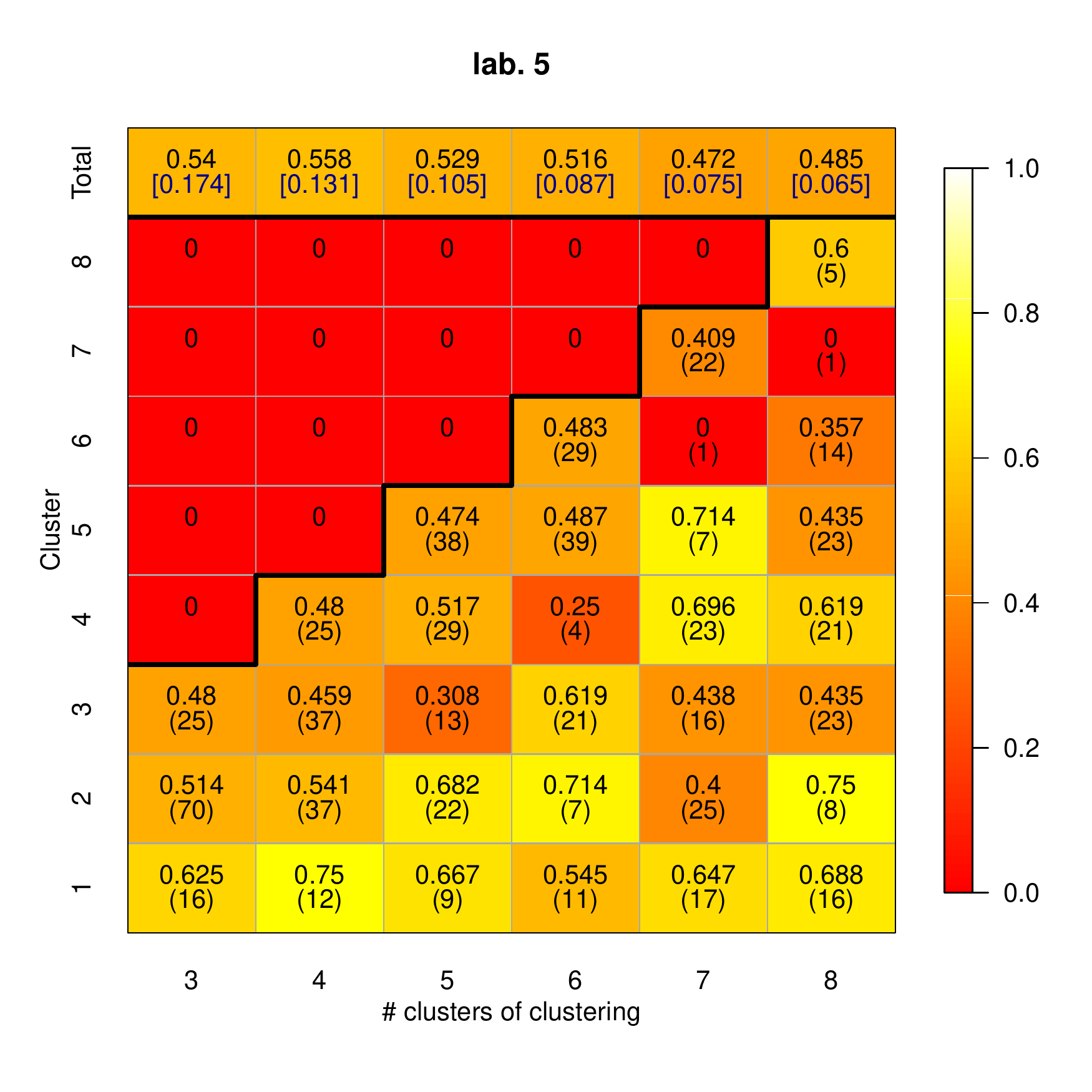}	\\
\end{tabular}
\end{center}
\caption{Example of ``heatmap" plots for a generic dataset and set of clusterings extracted using a given
clustering method for one set of \emph{labels} and one particular \emph{label} class distribution. 
Left: each cell represents a whole distinct clustering obtained using a given clustering algorithm, 
and the value $S_{\mathrm{tot}}$ and $S_{\mathrm{tot}}^{'}$ (in square brackets) of the total 
normal and weighted 
\emph{scores} respectively are reported for a set of \emph{labels}. Right: each cell (except for the cells 
in the upper row) represents a cluster and contains the cluster \emph{score} value and the number of 
members of the cluster (in parentheses). The upper row shows the total normal and weighted 
(in square brackets) \emph{score} values of each clustering.}  
\label{fig:heatmaps}
\end{figure*}

 For a given \emph{label} $L$ and a set of clusterings produced by the 
same clustering algorithm but with different number of clusters, the values of the \emph{scores} and the number of 
members of each cluster is shown in the ``heatmap" plot on the right side of Figure~\ref{fig:heatmaps}. 
This specific type of plot is useful to select large cluster \emph{score} values that may not be reflected 
in the global \emph{scores}, which are averaged over all clusters of the clustering (see 
equations~\eqref{eq:scoretot} and \eqref{eq:weightedscoretot}). The left side plot in Figure~\ref{fig:heatmaps} 
shows the values of the total normal and weighted \emph{scores} for distinct clusterings produced by a given 
clustering method as a function of the total number of clusters $N_{\mathrm{clust}}$, for a whole set 
of \emph{labels}. These plots can be used to determine whether multiple labels show similar trends in their 
degrees of correlation with the distribution of members of the clusters of each clustering. In this way,  
correlated attributes can be selected on the basis of the result of the clustering and labeling procedure and 
their dependences can be taken into account during the interpretation of the results. 

\subsection{Uncertainties}
\label{subsec:uncertainty}

The uncertainty on the \emph{features} of a dataset can affect the result of the clustering and 
the selection of correlations among the cluster distributions and the \emph{labels}. The 
clustering methods discussed in Sec.~\ref{subsec:clustering} do not take into account the presence of 
uncertainties on the attributes. The effect of the errors on the \emph{features} can be evaluated by applying 
CLaSPS to multiple realizations of the same dataset obtained by ``perturbing" the \emph{features} 
distribution and evaluating the spread of the \emph{scores} distribution relative to
the different clusterings. 

\begin{figure}
\begin{center}
\includegraphics[width=\linewidth]{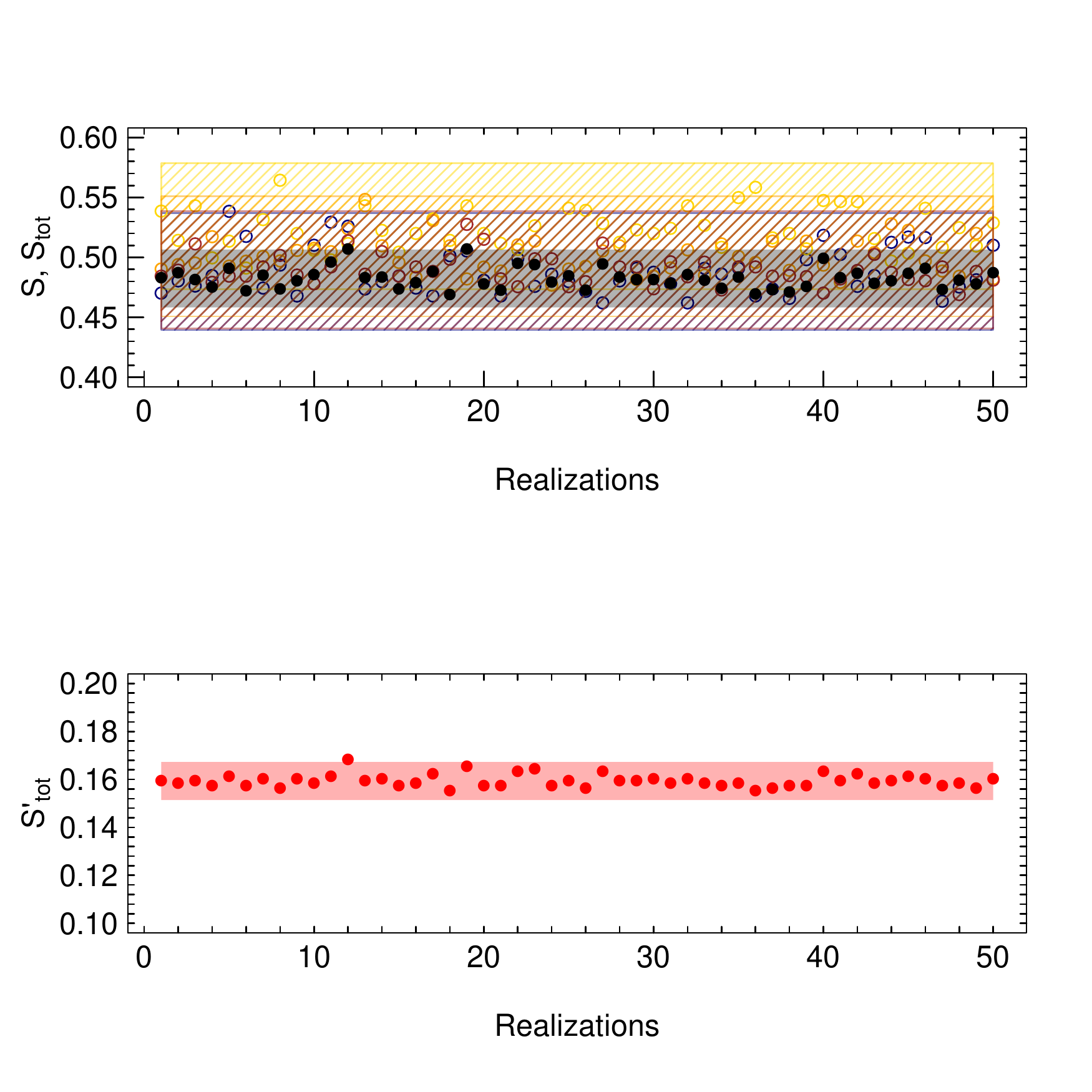}
\caption{Total \emph{scores} $S_{\mathrm{tot}}$ and $S^{'}_{\mathrm{tot}}$ distributions 
for 50 realizations of the \emph{features} of a generic dataset, as described in 
Sec.~\ref{subsec:uncertainty} (solid points). In the upper panel, the cluster and total 
normal \emph{scores} distributions for 4 distinct clusters and their clustering are 
plotted in full and open points respectively. In the lower panel, the weighted total \emph{scores} 
for the same clusterings are plotted. The $\pm 5\%$ and $\pm 10\%$ intervals around the 
mean values of the total and clusters \emph{scores} are drawn for reference in both panels. 
The variations of the cluster normal \emph{scores} can reach 10\% of their unperturbed values
(open circles in the upper panel), but the variations in the total normal and weighted (solid circles 
in upper and lower panels) \emph{scores} barely reach 5\% of their values. This fact shows the 
robustness of the total \emph{scores} values relatively to the presence of uncertainties on 
the \emph{features}.} 
\label{fig:scores_perturbations}
\end{center}
\end{figure}

In the case of the experiments described in this paper, multiple 
realizations of the dataset \emph{feature} distributions have been obtained by assuming that the error 
$\sigma_{\mathrm{x_{i}}}$ on the $i$-th \emph{feature} $x_{i}$ can be interpreted as the standard deviation 
of a normal error distribution. While this is a reasonable assumption for the uncertainties on the photometric 
quantities from large area surveys like the SDSS~\citep[e.g., ][]{fukugita1996} from which the \emph{features}
of the datasets discussed in this paper have been extracted, this method is general. 
For example, if the uncertainties on distinct \emph{features} have to be modeled with distinct 
distributions, the perturbations can be independently extracted for every single 
\emph{feature}, according to the same procedure described in the following for gaussian distributions. 
For the experiments discussed in this paper we
randomly extracted a distinct perturbing number $p_{i}$ from a gaussian distribution 
centered around zero and with
width equal to twice the uncertainty $\sigma_{x_{i}}$ on the value of the attribute $x_{i}$, for any 
given source of the dataset. The new realization of the $i$-th \emph{feature} $x_{i}$ is defined 
as follows:

\begin{equation}
x_{i}\!\rightarrow\!x^{'}_{i}\!=\!x_{i}\!+\!p_{i} 
\label{eq:pertubations}
\end{equation} 

\noindent where $p_{i}$ can be positive or negative: $p_{i}\!\in\![-\sigma_{\mathrm{x_{i}}},
\!\sigma_{\mathrm{x_{i}}}]$. This approach, in general, can be time-consuming as it requires
CLaSPS to be run multiple times on slightly different realizations of the same multi-dimensional
\emph{features} distribution. However, the high dimensionality of the \emph{feature} space 
where clustering methods are applied usually guarantees that the results of the clustering are robust. 
This statement can be verified by observing the distribution of total and cluster \emph{scores} 
values for several distinct realizations of the dataset obtained by perturbing the values of the 
\emph{features} as described above (see Figure~\ref{fig:scores_perturbations} for an example
of the distribution of the values of the clusters and total \emph{scores} distributions for 50 
realizations obtained with the above procedure). 

In the case of the datasets discussed in this paper and described in 
Sec.~\ref{sec:experiment1} and \ref{sec:experiment2}, arbitrary thresholds of 5\% and 10\% 
variation over the mean value of each total and cluster \emph{scores} respectively have been set to 
evaluate the stability of the clusterings. The results obtained confirmed that all clusterings are 
insensitive to the ``perturbations" to the values of the \emph{features} within these values of the 
threshold. This result was expected, since the datasets considered in this paper are sparse
in the \emph{feature} space where the clustering methods are applied, leading 
to intrinsically stable clustering configurations.

\section{Application to the CSC+ dataset}
\label{sec:experiment1}

The CSC+ is an example of the class of inhomogeneous datasets that have become common thanks
to the emerging VO technology. As discussed in Sec.~\ref{sec:introduction}, KD methods can 
improve the extraction of useful correlations from such samples, by minimizing the influence of biases and
selection effects inherent to federated data. In this specific case, we will show that the application of 
the CLaSPS method leads to the determination of simple well known relations between observables
from distinct spectral regions.

\subsection{The CSC+ dataset}
\label{subsec:dataset1}

CSC+ is a sample of spectroscopically selected optical quasars with X-ray observations in 
the Chandra Source Catalog~\citep{evans2010} (CSC) for which 
additional multi-wavelength photometric data are available. These sources have been classified as quasars 
using the Sloan Digital Sky Survey~\citep{aihara2011} (SDSS) spectroscopic observations. 
In addition, the sources of the CSC+ sample have 
been selected so that both near-infrared and ultraviolet photometric observations can be retrieved from the 
UKIRT Infrared Deep Sky Survey~\citep{lawrence2007} (UKIDSS) and the Galaxy Evolution 
Explorer~\citep{martin2005} (GALEX) catalogs respectively. 

The GALEX and UKIDSS counterparts to the sources and upper limits in both the CSC and SDSS 
surveys have been determined using pre-selected crossmatched catalogs containing all sources detected 
in the SDSS and in each of the two datasets discussed. More specifically, we have used the SDSS-GALEX 
crossmatched sample of sources~\citep{budavari2009} to determine the UV counterparts of the 
SDSS-Chandra sources, and the crossmatched table of the UKIDSS counterparts of the SDSS stellar 
sources for the IR photometry, available through the web interface to the GALEX database. 

The total number of sources of the CSC+ 
sample is 112 when considering only detection in the CSC (dataset used in experiments 1 and 2 
of Table~\ref{tab:featureslabels1}) and 192 including all sources with reliable Chandra upper limits 
for the flux in the Broad Chandra energy band as returned by the CSC sensitivity map service 
(dataset used in experiment 3). The final CSC+ sample is composed of radio-quiet quasars, except for
two sources that can be found in the VLA FIRST Survey Catalog~\citep{becker1995}.
More details on the specific data used to build the set of \emph{features} and \emph{labels} of the 
CSC+ sample described above and listed in Table~\ref{tab:featureslabels1} are discussed below.

CSC, the Chandra Source Catalog~\citep{evans2010} includes $\sim\!1.06\times10^5$ unique 
unresolved or slightly extended X-ray sources with 5-band photometry. The total cumulative sky coverage 
is 320 deg$^2$, but since the majority of sources have broad band fluxes of $\sim$10-14 cgs, the 
effective coverage is $\sim$260 deg$^2$. The 
sensitivity varies in different regions. A catalog containing CSC-SDSS positionally cross-matched 
sources\citep{evans2010} covers $\sim$133 deg$^2$, including $\sim$ 1.7$\times 10^{4}$ Chandra 
sources, of which $\sim$9000 have 
stellar and $\sim$7800 extended optical counterparts, mostly galaxies.

SDSS DR8~\citep{aihara2011} has observed $\sim\!1.4\times10^4$  deg$^2$ of the 
sky in 5 bands \emph{ugriz}, with photometric limiting magnitude of 22.2 in the \emph{r} band (95\% 
completeness for point sources). It includes spectra of $\sim\!1.8\!\times$10$^6$ sources in the 380-920 nm 
wavelength range. Classification in quasar, high-redshift quasars, galaxy, star and late-type 
stars classes and spectroscopic redshifts are available for these spectroscopically observed sources, 
based on the measured lines of the optical spectra. If emission lines are observed and if the source 
has a final redshift larger than 2.3,  it is classified as a high-redshift quasars.

UKIDSS~\citep{lawrence2007} has been designed to be the infrared counterpart to the SDSS and 
covers $\sim$7500 deg$^2$ of the sky in \emph{JHK} near-infrared bands to \emph{K}=18.3. The Large Area 
Survey (LAS) has imaged $\sim$ 4000 deg$^2$ (overlapping with the SDSS), with the additional \emph{Y} 
band to a limiting magnitude of 20.5. The final area of the overlap between CSC and UKIDSS LAS 
will be $\sim$50 deg$^2$. 

GALEX~\citep{martin2005} is a 2-band survey (far and near UV) that has observed the whole sky 
up to a limiting magnitude $\emph{nuv}\!=\!20.5$. It includes deep fields to magnitude 25 with 
spectroscopic observations.

\subsection{CSC+ dataset: \emph{features} and \emph{labels}}
\label{subsec:featureslabels1}

Three different experiments have been performed on the CSC+ sample using distinct combinations
of \emph{features} for the clustering and \emph{labels} for the evaluation of the \emph{scores}. The 
\emph{features} have all been extracted from the overall set of colors obtained from consecutive 
photometric filters, while the \emph{labels} are either photometric
measurements (not used for the clustering, in these cases) or the spectroscopic redshifts, classification flags 
and other parameters related to the shape of the spectral energy distributions of the sources. For example, 
the $\alpha_{OX}$ parameter~\citep{avni1982} has been used as a \emph{label}. 
The $\alpha_{OX}$ parameter 
measures the relative amount of energy emitted in the optical and X-rays and is defined as the 
spectral slope between the optical/UV and X-rays
monochromatic luminosities at $\mathrm{E}\!=\!2$\ KeV and $\lambda\!=\!2500$ \AA\ respectively:

\begin{equation} 
\alpha_{OX}\!=\!-\frac{\log{(\nu L_{\mathrm{\nu}})}_{\mathrm{Opt}}\!-\!\log{(\nu L_{\mathrm{\nu}}})_{\mathrm{X}}}
{\log{(\nu_{\mathrm{Opt}})}\!-\!\log{(\nu_{\mathrm{X}}})}\!+\!1
\end{equation}

\noindent The features of the experiments performed using the CSC+ sample are described in 
detail in Table~\ref{tab:featureslabels1}.  

\begin{table*}
\scriptsize
\begin{center}
\caption{List of the experiments performed on the CSC+ sample. A short description of dataset, 
the number of sources of the dataset, the total number of clusters for the clustering produced and the list of 
\emph{features} and \emph{labels} used are provided for each experiment. In the \emph{Labels} column, 
each \emph{label} is followed, in curly brackets, by the binning used to evaluated the 
\emph{scores}. For categorial \emph{labels}, the binning is specified by providing the actual values 
corresponding to the distinct classes; for continuous \emph{labels}, the extremes of the bins defining 
the classes are provided.}
\begin{tabular}{llccll}
\tableline
\tableline
Experiment	& Dataset				& \# sources	&\# clusters	&	\emph{Features}			& \emph{Labels}													\\
\tableline
Exp. 1		& SDSS quasars with	&	112		&	\{3, 4, 5\}	&$fuv$-$nuv$,$nuv$-$u$,$u$-$g$,	& $z_{\mathrm{spec}}$\{0.4,1.1,1.9\},$HR(ms)$\{-0.4,-0.2,0\},					\\
			& CSC detection		&			&			&$g$-$r$,$r$-$i$,$i$-$z$,$z$-$Y$,	& $HR(hm)$\{-0.1,0,0.2\},												\\
			& 					&			&			&$Y$-$J$,$J$-$H$,$H$-$K$		& $L(B)$\{2,4,6,8\}$\times 10^{43}$ erg s$^{-1}$,							\\	
			&					&			&			&							& $\alpha_{\mathrm{OX}}$\{1.3\}										\\
Exp. 2		& SDSS quasars with	&	112		&	\{3, 4, 5\}	&$fuv$-$nuv$,$nuv$-$u$,$u$-$g$, 	& $z_{\mathrm{spec}}$\{0.4,1.1,1.9\},		 							\\
			& CSC detections		&			&			&$g$-$r$,$r$-$i$,$i$-$z$,$z$-$Y$,	& $\alpha_{\mathrm{OX}}$\{1.3\},										\\
			& 					&			&			&$Y$-$J$,$J$-$H$,$H$-$K$,		& $L(B)$\{2,4,6,8\}$\times 10^{43}$ erg s$^{-1}$							\\
			&					&			&			&$HR(ms)$,$HR(hm)$			&																\\
Exp. 3		& SDSS quasars with	&	192		&	\{3, 4, 5\}	&$fuv$-$nuv$,$nuv$-$u$,$u$-$g$, 	& $z_{\mathrm{spec}}$\{0.4,1.1,1.9\},	 								\\
			& CSC detections and 	&			&			&$g$-$r$,$r$-$i$,$i$-$z$,$z$-$Y$,	& $\alpha_{\mathrm{OX}}$\{1.3\},$f_{\mathrm{Xdet}}$\{0,1\} 					\\
			& upper limits			&			&			&$Y$-$J$,$J$-$H$,$H$-$K$		& $L(B)$\{2,4,6,8\}$\times 10^{43}$ erg s$^{-1}$							\\
\tableline
\label{tab:featureslabels1}
\end{tabular}
\end{center}
\end{table*}

\subsection{Results of the application of CLaSPS to CSC+}
\label{subsec:application1}

\begin{figure*}
\begin{center}
\begin{tabular}{ccc}
\includegraphics[width=0.3\linewidth]{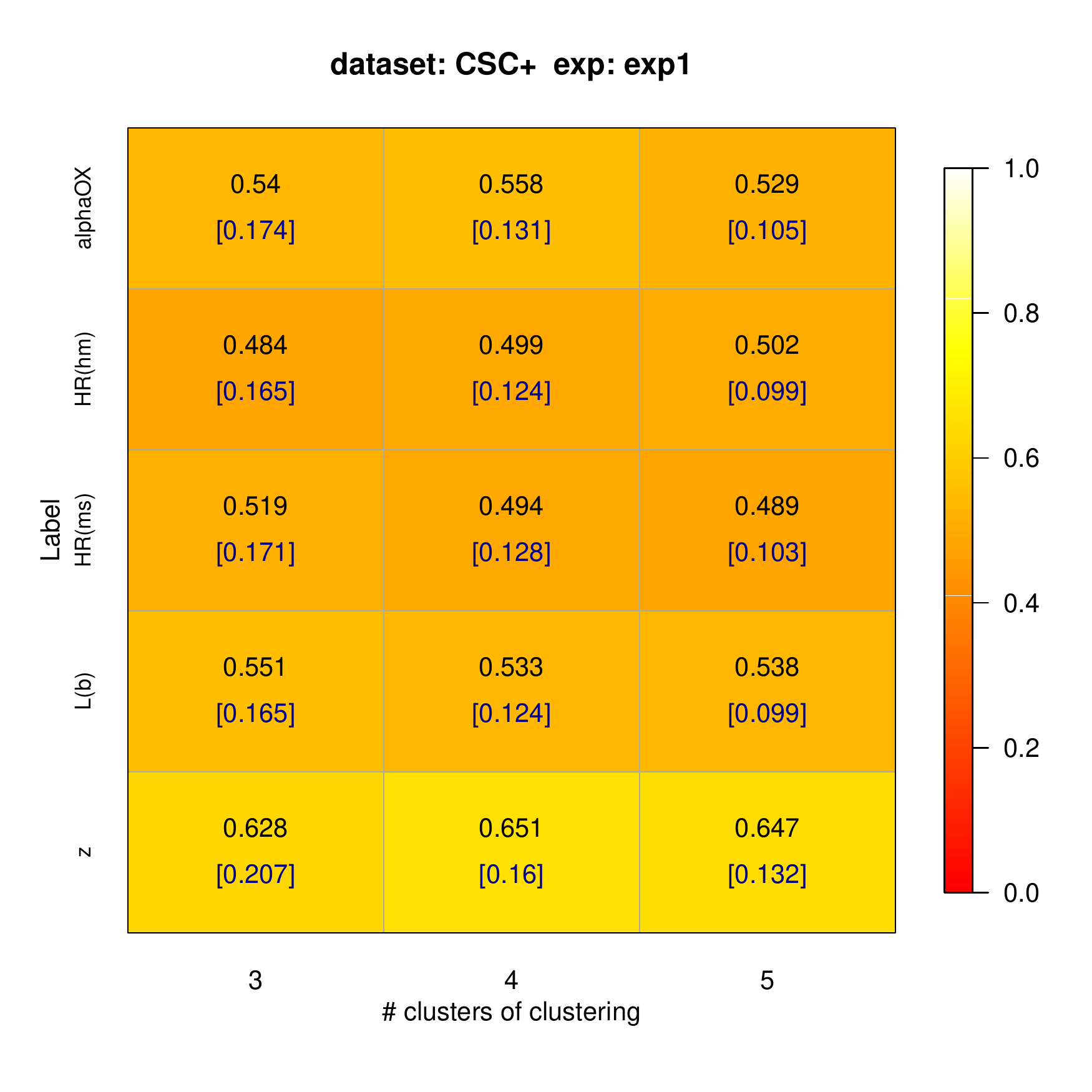}	&
\includegraphics[width=0.3\linewidth]{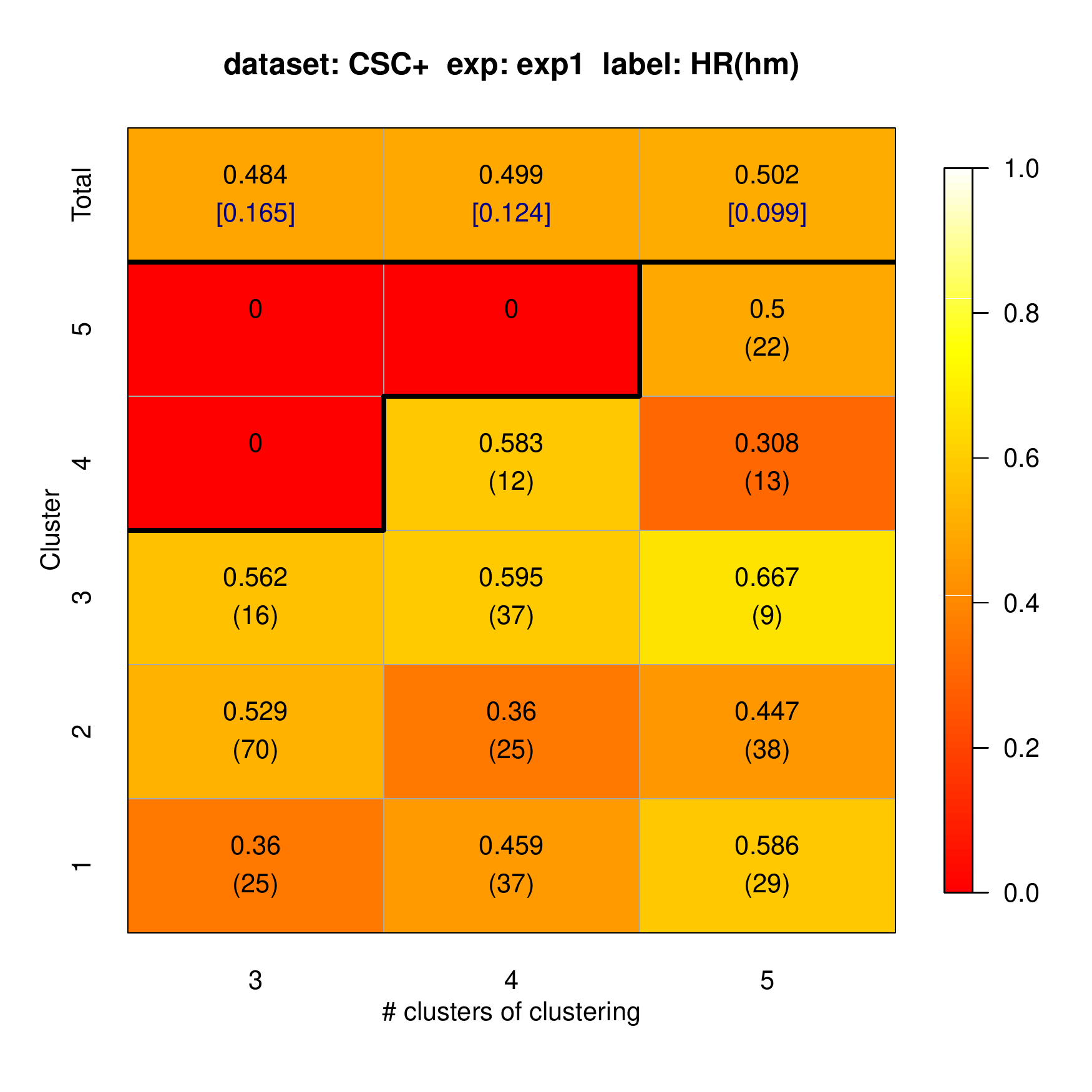}	&
\includegraphics[width=0.3\linewidth]{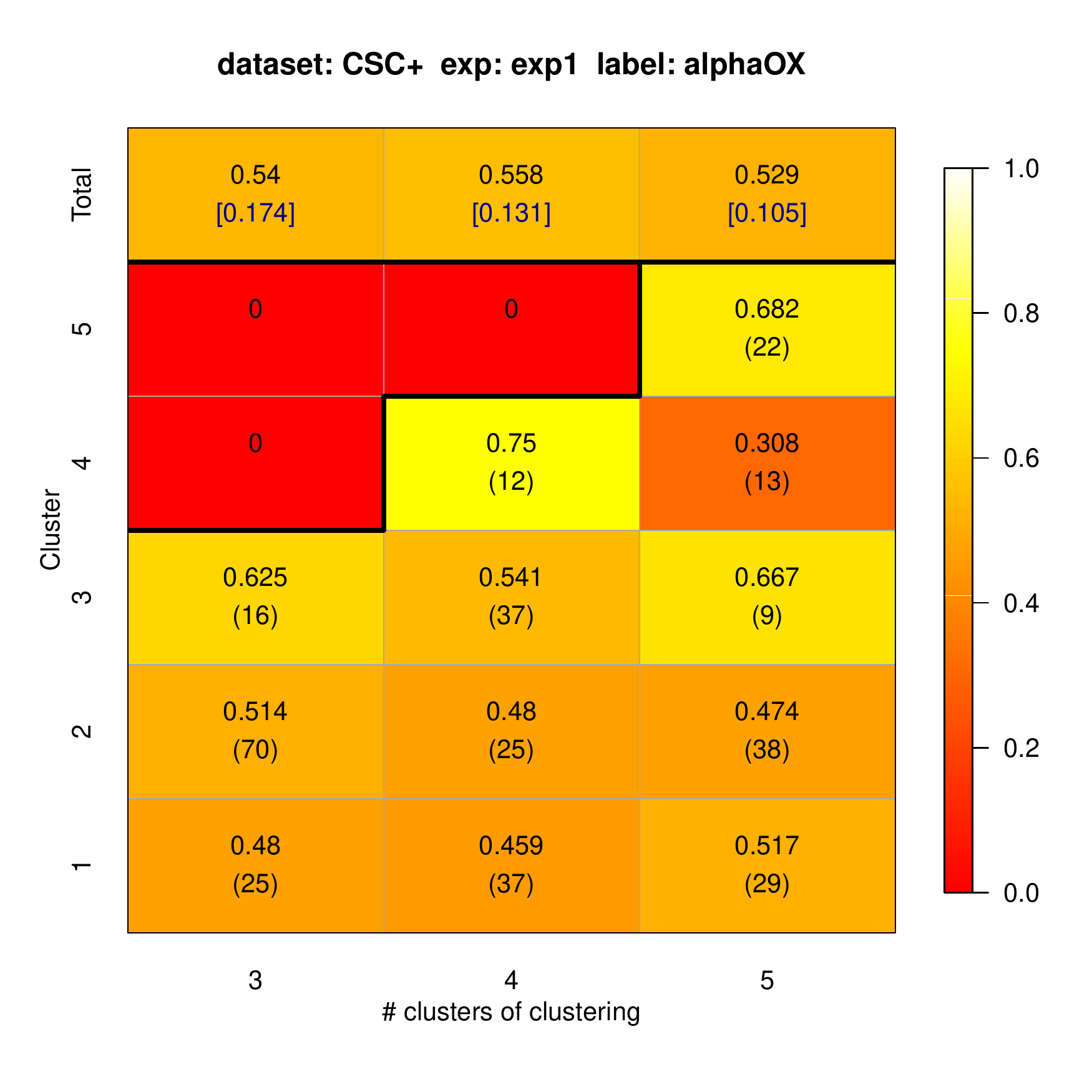}	\\
\includegraphics[width=0.3\linewidth]{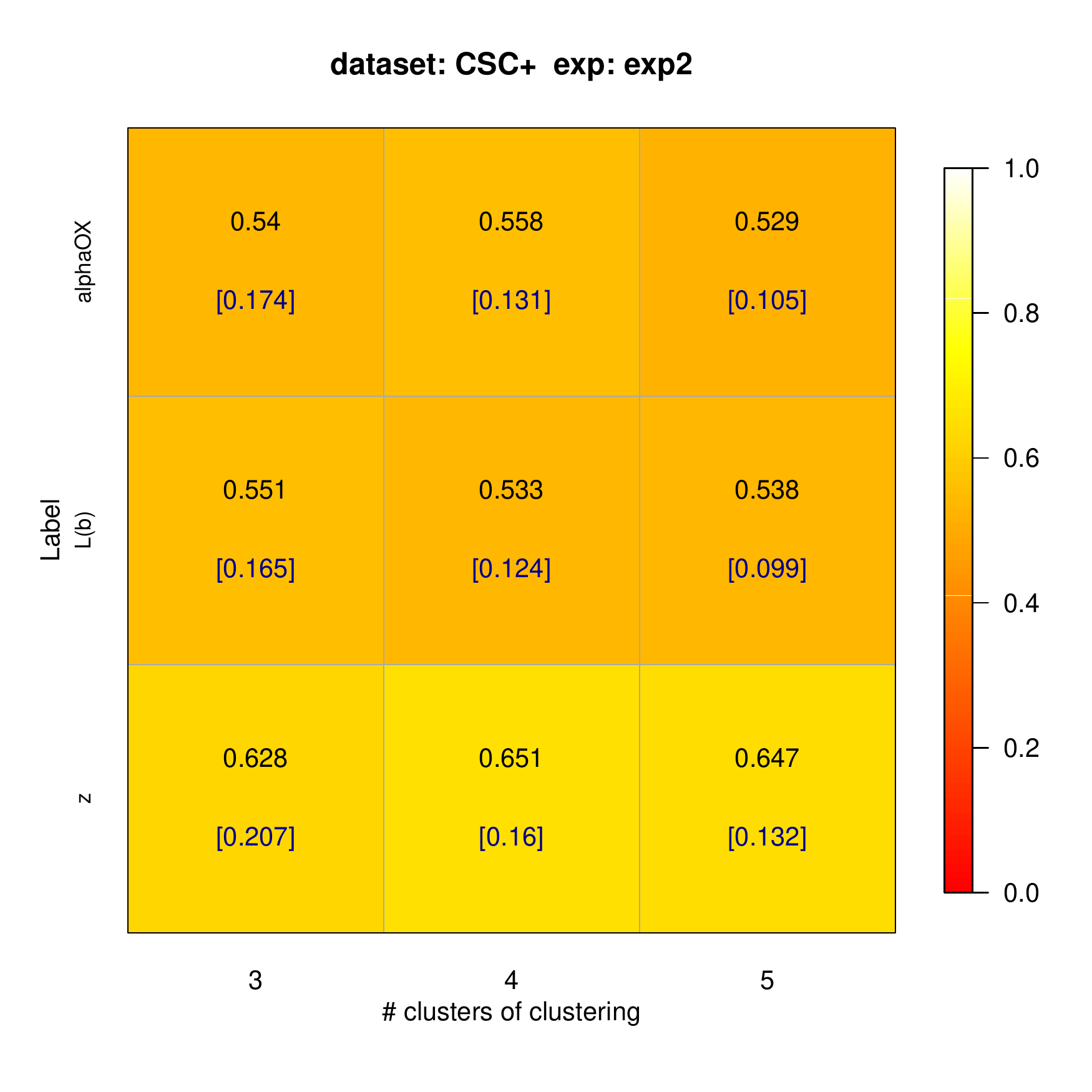}	&
\includegraphics[width=0.3\linewidth]{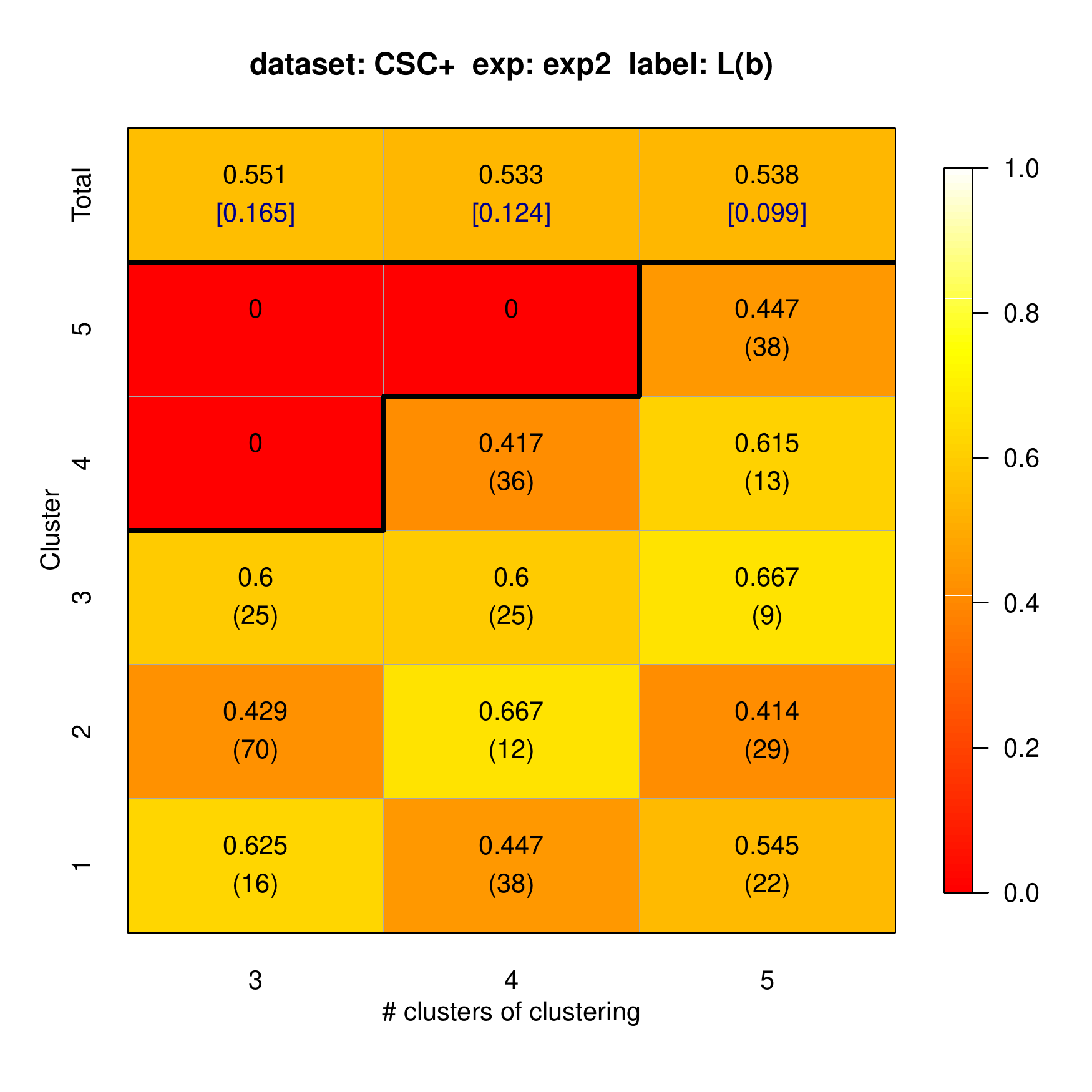}	&
\includegraphics[width=0.3\linewidth]{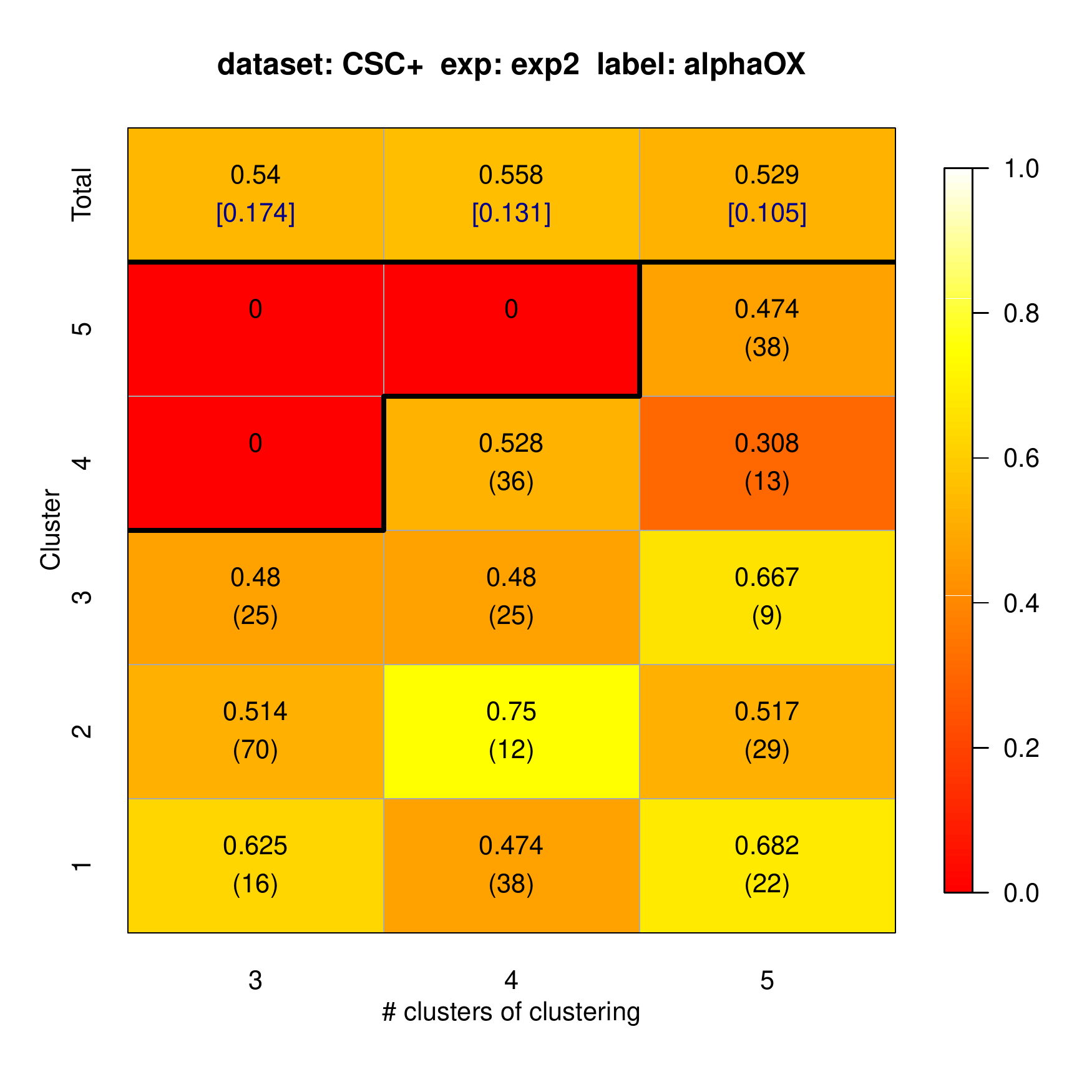}	\\
\includegraphics[width=0.3\linewidth]{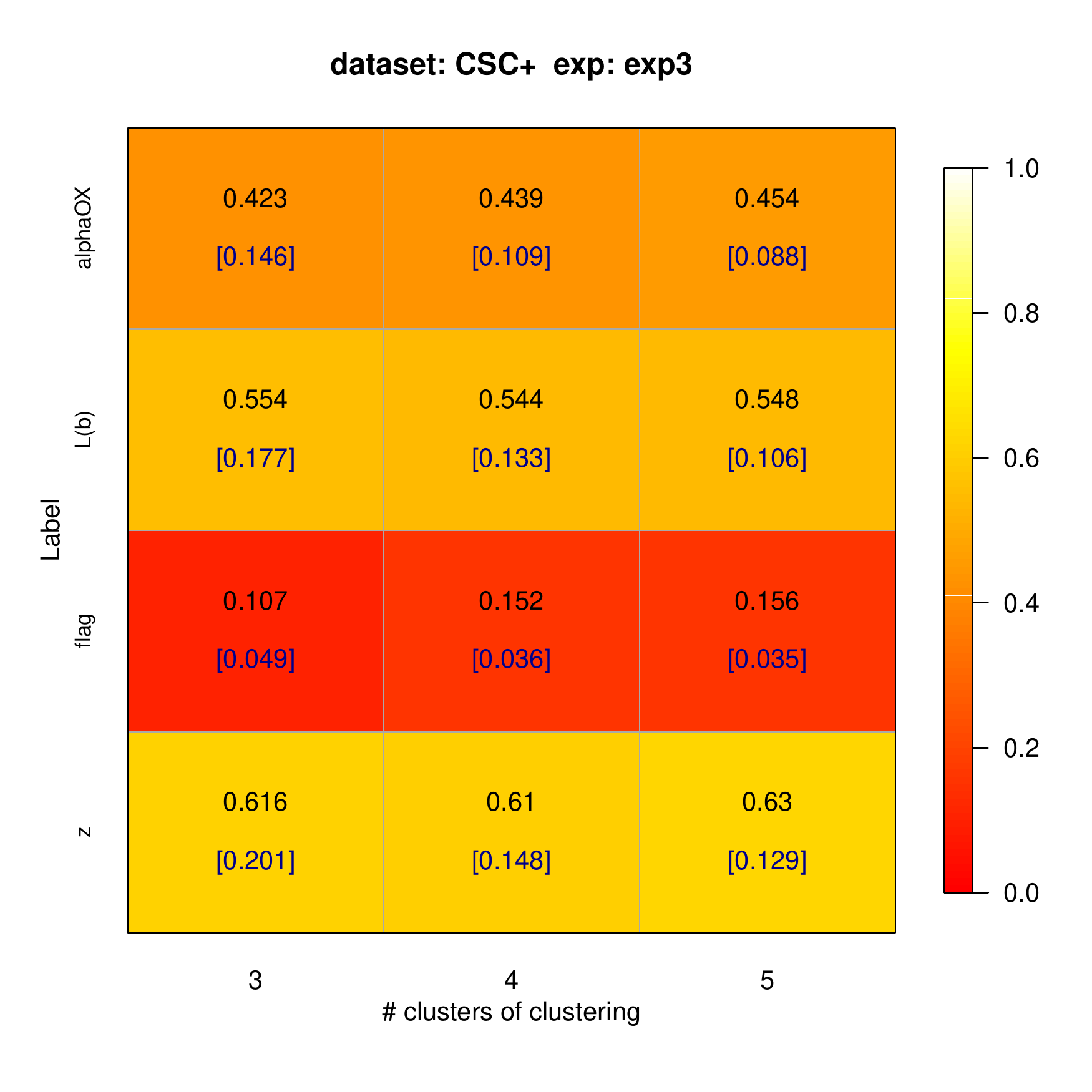}	&
\includegraphics[width=0.3\linewidth]{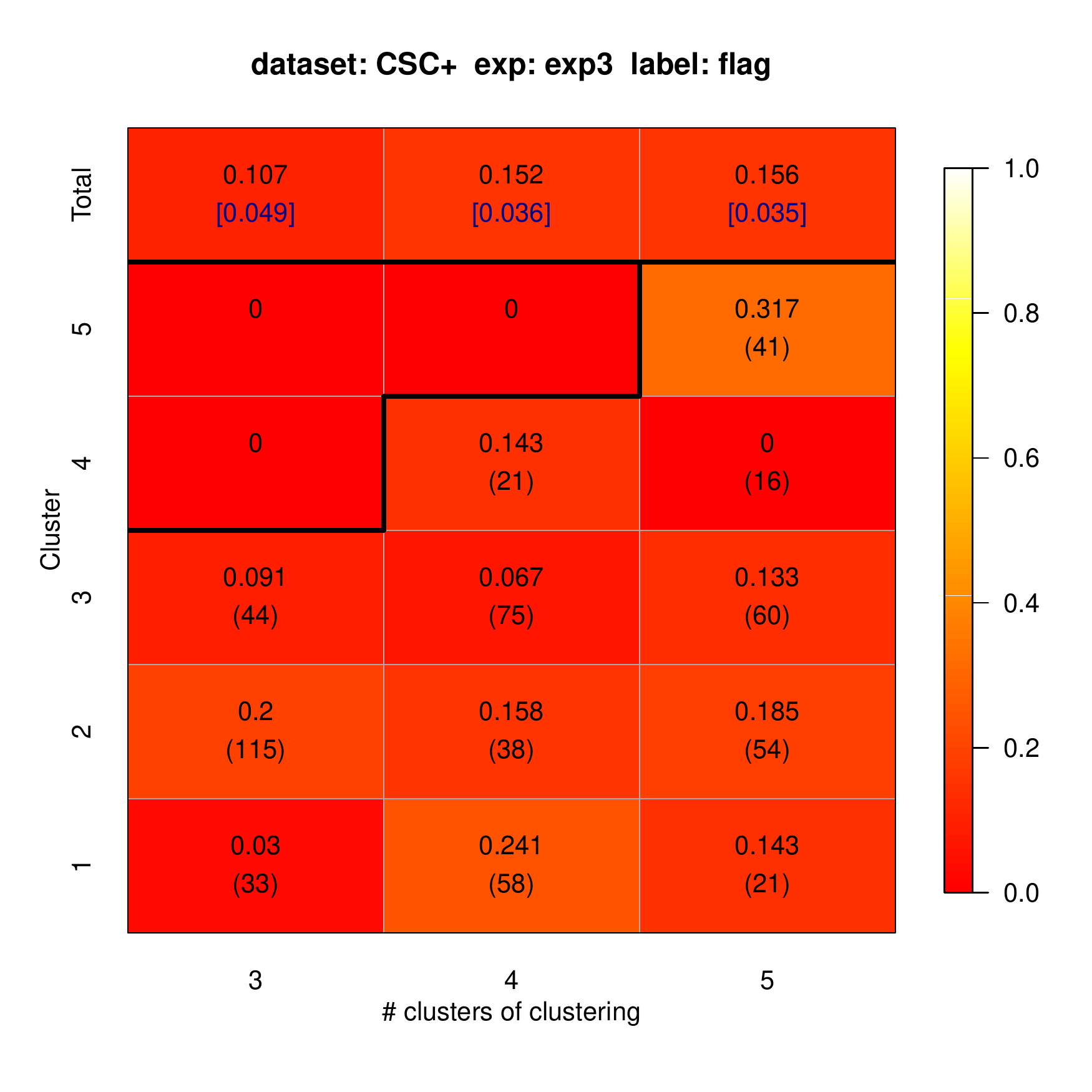}	&
\includegraphics[width=0.3\linewidth]{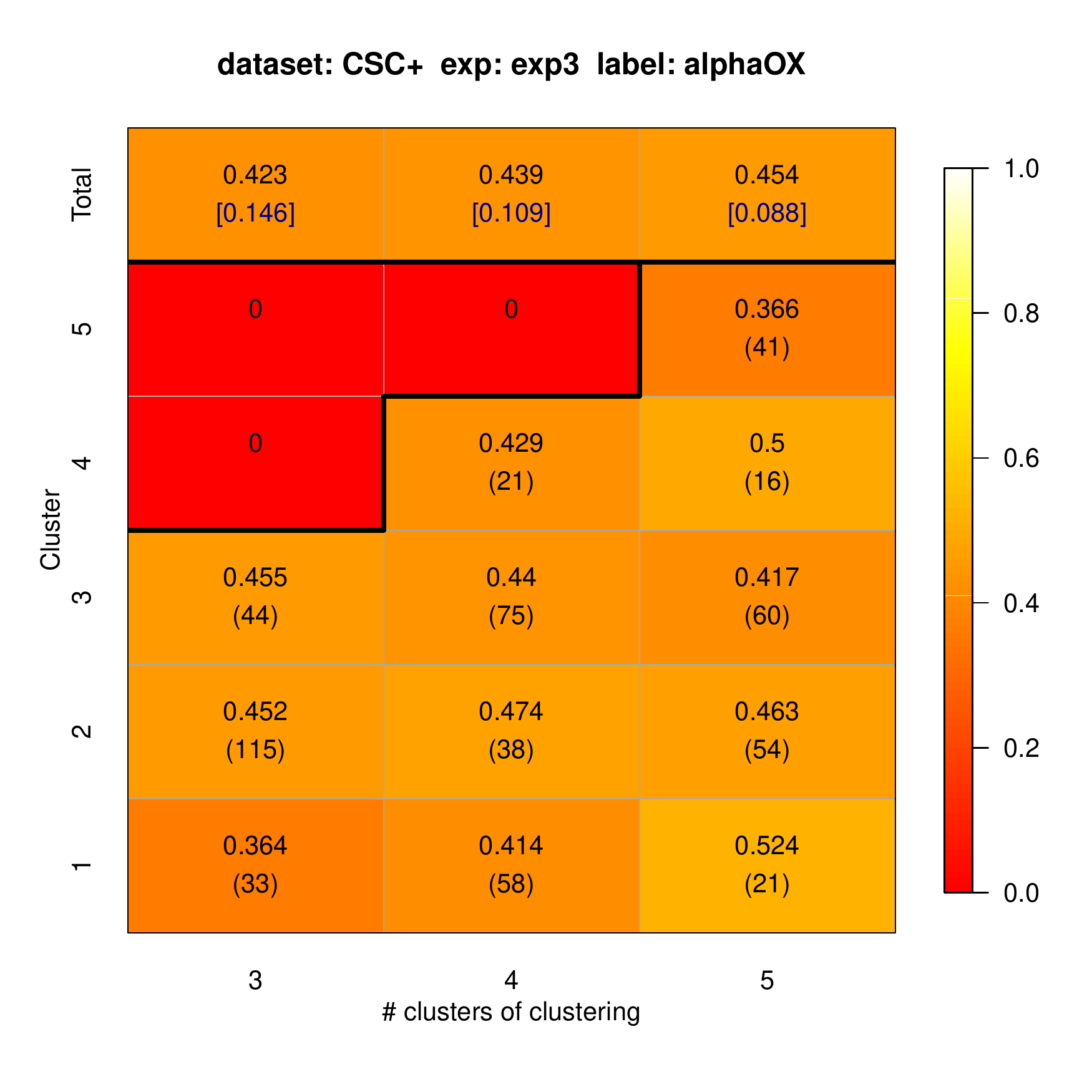}	\\
\end{tabular}
\end{center}
\caption{Results of the application of the CLaSPS method to the datasets based on the CSC+ sample (for details
see Sec.~\ref{sec:experiment1}). In the upper row, plots from the Exp.1 are shown. From left to right: 
total \emph{scores} distribution for the clusterings obtained with the K-means method; cluster \emph{scores} 
distribution for the \emph{label} X-ray hardness ratio $HR(hm)$; cluster \emph{scores} distribution for the 
$\alpha_{\mathrm{OX}}$ index. The right plot, in particular, has been discussed extensively in 
Sec.~\ref{subsec:application1}. In the central row,
plots from the results of Exp.2 are showed. From left to right: total \emph{scores} distribution for the 
clusterings obtained with the K-means method; 
cluster \emph{score} distribution for the total X-ray luminosity used as \emph{label}; cluster 
\emph{scores} distribution for the $\alpha_{\mathrm{OX}}$ index. 
In the lower row, the plots from the Exp.3 are shown. From left to right: total \emph{score} distributions 
of the clusterings obtained using the SOM method; cluster \emph{scores} distribution for the \emph{label} 
represented by the X-ray detection flag; cluster \emph{scores} distribution for the $\alpha_{\mathrm{OX}}$ 
index (for more details, see Sec.~\ref{tab:featureslabels1}). The discussion in Sec.~\ref{subsec:application1} 
and Figure~\ref{fig:alphaOX} are based on the 
clustering with five total clusters of the second experiment (mid-left plot in this figure), 
using as \emph{label} the $\alpha_{\mathrm{OX}}$
spectral index.}  
\label{fig:CSCexps}
\end{figure*}

\begin{figure}
\begin{center}
\includegraphics[width=\linewidth]{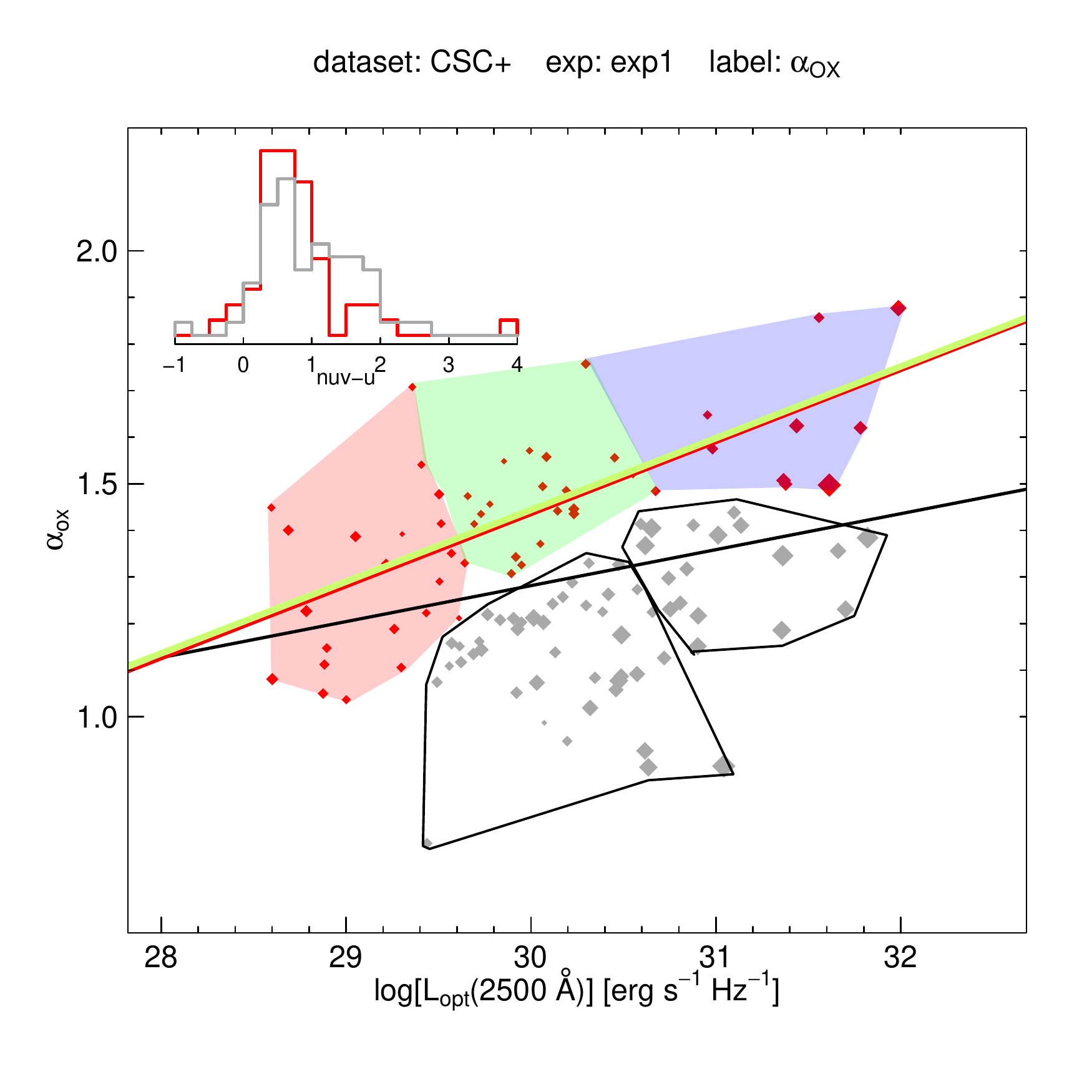}
\end{center}
\caption{Distribution of the sources in the first experiment (Exp.~1) with the CSC+ sample in the 
$L_{\mathrm{opt}}(2500\AA)$ vs $\alpha_{\mathrm{OX}}$ plane. The shaded regions 
containing the red symbols
correspond to the projections of the three clusters (clusters one, three and five shown in
the upper-right plot in Figure~\ref{fig:CSCexps}) with large \emph{scores} values for 
the label $\alpha_{\mathrm{OX}}$
in the first experiment (Exp. 1 in Table~\ref{tab:featureslabels1}). The two black polygons
represent the projections of the remaining two clusters (clusters two and four in the mid-left heatmap
in Figure~\ref{fig:CSCexps}) with small \emph{scores} values.
The size of the symbols used for the plots is proportional to the near-ultraviolet color $nuv-u$ of 
each source, and in the inset of the plot the histograms of the $nuv-u$ color distribution for
sources belonging to the three interesting clusters and the two unselected clusters is shown. 
The red and black lines represent the linear regression for the points in the correlated clusters 
and the whole sample, respectively. The red line is in perfect
agreement with the best-fit relation from~\citep{lusso2010} 
(green line), derived from a sample of 545 AGNs.}
\label{fig:alphaOX}
\end{figure}

The main results of the application of CLaSPS to the two different datasets based on the CSC+ 
sample are summarized in the plots in Figures~\ref{fig:alphaOX} and~\ref{fig:CSCexps}.

In the first experiment (see Table~\ref{tab:featureslabels1}), we have found a significant 
correlation between the near-infrared, optical and 
ultraviolet colors, used as \emph{features}, and the $\alpha_{\mathrm{OX}}$ index for two clusterings, 
composed of 4 and 5 clusters respectively and produced using the K-means 
and SOM methods (see leftmost plot in the upper row of Figure~\ref{fig:CSCexps}).  
Even if the total normal and weighted \emph{scores} values for the clustering 
composed of 5 total clusters are smaller than the scores for the other two clusterings, 
this clustering has been considered more interesting because of the larger number 
of sources contained in clusters with significantly large values of the cluster \emph{scores}. 
We will discuss here the correlation involving the members of the second, third and fourth 
clusters of the clustering composed of five total clusters (see upper-left plot in Figure~\ref{fig:CSCexps}). 

In order to determine whether there is a subset of \emph{features} responsible for 
the correlation observed, a Principal Component Analysis~\citep{hartigan1975} (PCA) has been 
performed on the \emph{feature} distribution of these three clusters. 
The PCA finds that the correlation is mostly due to the blue optical/UV
colors $nuv-u$ and $u-g$. 

The correlation between the optical blue and near-UV features of quasars 
and the $\alpha_{\mathrm{OX}}$ spectral index has been discussed in several papers  
in the literature~\citep[see, for example][]{vignali2003,lusso2010}, and can be explained on the 
basis of the definition of the $\alpha_{\mathrm{OX}}$ spectral index itself (given in 
Sec.~\ref{subsec:featureslabels1}), and the characteristics of the spectral energy distributions 
(SED) of homogeneous samples of radio-quiet quasars observed in the X-rays. In particular, 
the presence of this correlation is usually associated with the presence of a prominent 
component of the SED of the quasars at near-UV wavelengths, called ``big-blue-bump".

In Figure~\ref{fig:alphaOX}, we show the $L_{\mathrm{opt}}(2500\AA)$ vs
$\alpha_{\mathrm{OX}}$ distribution of the sample used in the first experiment for the clustering composed
of five clusters. The projections of the five clusters are plotted as shaded colored regions and closed 
black lines for the three clusters with large \emph{score} values (clusters one, three and five in 
the upper-right plot in Figure~\ref{fig:CSCexps}) and the remaining two clusters (clusters two and 
four in the upper-right plot in Figure~\ref{fig:CSCexps}) respectively.
 
The clusters with large \emph{scores} are those for which the correlation between the optical mono-chromatic 
luminosity and the $\alpha_{\mathrm{OX}}$ index is more significant. In fact, 
while the significance of finding a correlation between the two parameters for 
the whole sample is low ($\sim 43\%$), the significance is
larger for the subset of points of the three clusters selected ($\sim 90\%$). In Figure~\ref{fig:alphaOX},
the size of the symbols are proportional to $nuv-u$ color values of the sources, and it is evident that 
the members of the correlated clusters  have, on average, lower values of $nuv-u$. This suggests that 
the SEDs of the sources belonging to the three clusters with large \emph{scores} values are dominated 
by the ``big blue bump'' component (the red and black lines are associated with the best 
fitting linear relations for the members of the three clusters and the total sample respectively, and are 
shown only for reference). In this regard, we conclude that, despite the fact that the CSC+ sample 
used for this experiment is highly inhomogeneous, CLaSPS select a subset of clusters whose members show a 
significant degree of correlation between the optical mono-chromatic luminosity $L_{\mathrm{opt}}(2500\AA)$ 
and the $\alpha_{\mathrm{OX}}$ spectral index. The behavior of this subset of sources, is in agreement 
with the results found and discussed in literature for homogeneously drawn samples of X-ray emitting
radio-quiet optically selected quasars. 

A similar correlation pattern, weaker than the one observed for the 
$\alpha_{\mathrm{OX}}$ index though, is observed in the distribution 
of \emph{score} values for the hardness ratio $HR(hm)$ used as 
\emph{label} (see central plot of the upper row in Figure~\ref{fig:CSCexps}). 

In the second experiment (Exp.~2 in Table~\ref{tab:featureslabels1}), the hardness ratios $HR(hm)$ 
and $HR(ms)$ have been used as
\emph{features}, together with the other variables used as \emph{features} in the first experiment. 
Correlations similar to those observed in the first experiment are observed 
for in the clusterings with 4 and 5 total clusters each in the second experiment 
(see central and left plots of the central row in Figure~\ref{fig:CSCexps}).
 
The PCA of the \emph{feature} distribution of the clusters with larger values of the 
\emph{scores} showed that the correlation can be mostly attributed to a subset of 
\emph{features} including the $nuv-u$ and $u-g$ colors and both the X-rays hardness ratios. 

In the third experiment (Exp.~3 in Table~\ref{tab:featureslabels1}), we have used as \emph{label} 
the values of upper limits for X-ray luminosity and considered them as detections in order to test 
whether the distribution of not-X-ray colors of the sample correlates with either the detections or 
the upper limits observed in X-rays. As shown in the plots of the lower row in 
Figure~\ref{fig:CSCexps}, no interesting correlations among the 
set of \emph{features} and the \emph{labels} considered is visible. 
In particular, the X-ray detection flag is not 
correlated with the near-infrared, optical and ultraviolet colors used as \emph{features},
regardless of the clustering methods and total number of clusters of the clusterings.

\section{Application to the Blazars dataset}
\label{sec:experiment2}

CLaSPS has also been applied to
a sample of blazars with available photometric data spanning from the mid-infrared to optical
wavelengths, with
additional information in the $\gamma$-ray spectral range. Blazars are a peculiar family of AGNs characterized 
by rapid variability at all frequencies. The other distinguishing observational properties of blazars include 
flat radio spectra, high observed luminosity and highly variable radio to optical polarization. 
Blazars are a dominant class of extragalactic sources at radio, microwave and $\gamma$-ray frequencies. The 
observational characterization of this class of galaxies is interesting as a tool to shed some light on the physical
mechanisms responsible for the emission. The experiments described have aimed at the characterization of the 
blazars population in the infrared bands, extending the type of analysis already performed on 
2MASS~\citep{skrutskie2006} data~\citep[e.g.][]{chen2005} to longer wavelengths, using the recently 
released WISE~\citep{wright2010} mid-infrared photometric data.

\subsection{The Blazars dataset}
\label{subsec:dataset2}

The Blazars sample is based on the ROMA-BZCAT~\citep{massaro2009} list of blazars. This catalog assembles
blazars known in the literature and confirmed by the inspection of their multi-wavelength 
emission. The members of the ROMA-BZCAT catalog are selected on the basis of a set of criteria 
involving the presence of detection in the radio band  down to 1 mJy flux density at 1.4 GHz (2.1 $\mu m$), 
the optical identification and availability of an optical spectrum for further spectral classification and the detection 
of isotropic X-ray luminosity $L_{X} \geq 10^{43}$ erg $\times s^{-1}$. Such criteria
do not yield a statistically homogeneous or complete sample of blazars but provide the largest and more 
carefully selected sample of confirmed blazars available to date. 
In the ROMA-BZCAT, blazars are also divided in three spectral classes, based on the prominence of the emission 
features in the optical spectra of these sources: BZB for the BL Lac sources, i.e. 
AGNs with featureless optical spectra and narrow emission lines; BZQ for flat-spectrum radio quasars 
with optical spectra showing broad emission lines and typical blazars behavior; BZU for blazars of 
uncertain type, associated with sources with peculiar characteristics but also showing typical traits of 
the blazars (a more detailed description of on the Blazars sample can be found 
in~\citep{massaro2011,dabrusco2012}). This sample includes $\sim800\ \gamma$-ray 
sources from the 2FLG~\cite{abdo2011} associated with members of the catalog to a 
high level of confidence, and 571 of these blazars 
are also present in the ROMA-BZCAT. More details on the specific data used as \emph{features} 
and \emph{labels} of the Blazars sample can be found below (for SDSS data, 
see Sec.~\ref{subsec:dataset1}).

2MASS~\citep{skrutskie2006} has uniformly scanned the whole near-infrared sky in three
bands $H$, $J$ and $K_{s}$ detecting points sources brighter than $\sim 1$ mJy in each filter, with
with positional accuracy of $0.4^{\prime\prime}$, to a magnitude limit (for stellar sources 
in unconfused regions and outside of the galactic plane) of 15.8 in the $J$ band;
	
The WISE mission~\citep{wright2010} has observed the entire sky in the mid-infrared spectral 
interval at 3.4, 4.6, 12, and 22 $\mu$m with an angular resolution of 6.1$^{\prime\prime}$, 
6.4$^{\prime\prime}$, 6.5$^{\prime\prime}$ \& 12.0$^{\prime\prime}$ in the four bands, achieving 
5$\sigma$ point source sensitivities of 0.08, 0.11, 1 and 6 mJy in 
unconfused regions on the ecliptic, respectively. The astrometric accuracy of WISE
is $\sim 0.50^{\prime\prime}, 0.26^{\prime\prime}, 0.26^{\prime\prime}$, and 1.4$^{\prime\prime}$ 
for the four WISE bands respectively.

The 2FLG catalog~\citep{abdo2011} contains primarily unresolved sources detected in the 
all-sky Fermi observations obtained throughout the second year of operation. The sources, 
after detection and the localization in the sky, are assigned, among other parameters,
an integrated flux in the 100 MeV to 100 GeV energy range, a spectral shape and a significance 
parameter {\it TS} based on how significantly each source emerges from the background. Only 
sources with {\it TS $\geq$ 25}, corresponding to a significance of $4\sigma$, have been included in 
the catalog. Each of the 1873 2FLG sources have been considered for identification with 
already known astronomical sources available in 
literature multi-wavelength observations. For 127 of the 2FLG  sources firm identifications have 
been produced (namely, reliable identifications based on synchronous periodic variability of the sources, 
coincident spatial morphologies for extended sources or correlated aperiodic variability). The remaining
sources have been investigated for association with sources contained in a list of source catalogs based on 
different multi-wavelength observations. The BZCAT~\citep{massaro2009} catalog is one of the 
catalogs used for the association of the 2FLG sources, and 571 2FGL sources have been associated with
a BZCAT blazar. The $\gamma$-ray detection flags used in the experiments described in this paper
are based on the official associations of the 2FLG sources from~\citep{abdo2011}.

\subsection{Blazars dataset: features and labels}
\label{subsec:featureslabels2}

In the first experiment, we have used as \emph{features} the 
colors calculated with consecutive filters from mid-infrared (WISE) to the optical (SDSS). As
\emph{labels}, we have used the spectroscopic classification in BZB, BZQ and BZU from the 
ROMA-BZCAT catalog~\citep{massaro2009}, 
the radio flux density at the frequency $\nu = 1.4$ GHz and a $\gamma$-ray detection flag. Such flag
is equal to 1 for the ROMA-BZCAT sources than have been associated with a $\gamma$-ray source 
from 2FGL catalog~\citep{abdo2011}, and 0 for all the other sources. 

In the second experiment, the \emph{labels} used for the first experiment Exp.~1 in 
Table~\ref{tab:featureslabels2} have been complemented by WISE colors, not
used as \emph{features} in this case, while the Fermi detection flags have not been used
as \emph{label} because this sample is composed of all the blazars of our sample associated with Fermi 
detections. 
In the third experiment, only the optical and near-infrared
colors from SDSS and UKIDSS respectively have been used as \emph{features}, while the WISE mid-infrared
colors and the Fermi detection flag have
been added to the set of \emph{labels} already used in the first experiment for all
blazars, regardless of their association with Fermi detections.
The parameters of the two experiments are shown in details in Table~\ref{tab:featureslabels2}.

\begin{table*}
\scriptsize
\begin{center}
\caption{Characteristics of the experiments performed on the Blazars sample. For a detailed description of the 
columns, refer to the caption of the Table~\ref{tab:featureslabels2}.}
\begin{tabular}{llccll}
\tableline
\tableline
Experiment	& Dataset				&\# sources			&\# clusters	&\emph{Features}					& \emph{Labels}		\\
\tableline
Exp. 1		& BZCAT blazars with 	& 241				&\{3, 4, 5, 6\}	&$u$-$g$,$g$-$r$,$r$-$i$, 			& source class\{BZB, BZQ, BZU\},						\\
			& FIR, NIR and Optical 	&					&			&$i$-$z$,$z$-$J$,$J$-$H$,			&$f(1.4\mathrm{GHz})$\{10$^{3}$,3$\times 10^{3}$\}			\\
			& photometry			&					&			&$H$-$K$,$K$-$[3.4]$,$[3.4]$-$[4.6]$,	&												\\
			& 					&					&			&$[4.6]$-$[12]$,$[12]$-$[22]$			&												\\						
Exp. 2		& $\gamma$-ray detected& 241				&\{3, 4, 5, 6\}	&$u$-$g$,$g$-$r$,$r$-$i$,			& source class\{BZB, BZQ, BZU\},						\\
			& BZCAT blazars with	&					& 			&$i$-$z$,$z$-$J$,$J$-$H$,			&$f(1.4\mathrm{GHz})$\{10$^{3}$,3$\times 10^{3}$\},			\\	
			& FIR, NIR and Optical	&					&			&$H$-$K$,$K$-$[3.4]$,$[3.4]$-$[4.6]$,	& 												\\
			& photometry	 		&					&			&								& 												\\			
Exp. 3		& BZCAT blazars with	& 241				&\{3, 4, 5, 6\}	&$u$-$g$,$g$-$r$,$r$-$i$,			& source class\{BZB, BZQ, BZU\},						\\
			& NIR and Optical		&					&			&$i$-$z$,$z$-$J$,$J$-$H$,			& $f(1.4\mathrm{GHz})$\{10$^{3}$,3$\times 10^{3}$\},		\\
			& photometry			&					&			&$H$-$K$						& $f_{\mathrm{\gamma det}}$\{0,1\},$[3.4]$-$[4.6]$\{0,0.5,1,	\\ 
			&					&					&			&								& 1.5,2\},$[4.6]$-$[12]$\{0,1,2,3,4,5\},					\\			
			& 					&					&			&								& $[12]$-$[22]$\{0,1,2,3,4\}							\\				
\tableline
\label{tab:featureslabels2}
\end{tabular}
\end{center}
\end{table*}

\subsection{Results of the application of CLaSPS to Blazars dataset}
\label{subsec:application2}

\begin{figure*}
\begin{center}
\begin{tabular}{ccc}
\includegraphics[width=0.33\linewidth]{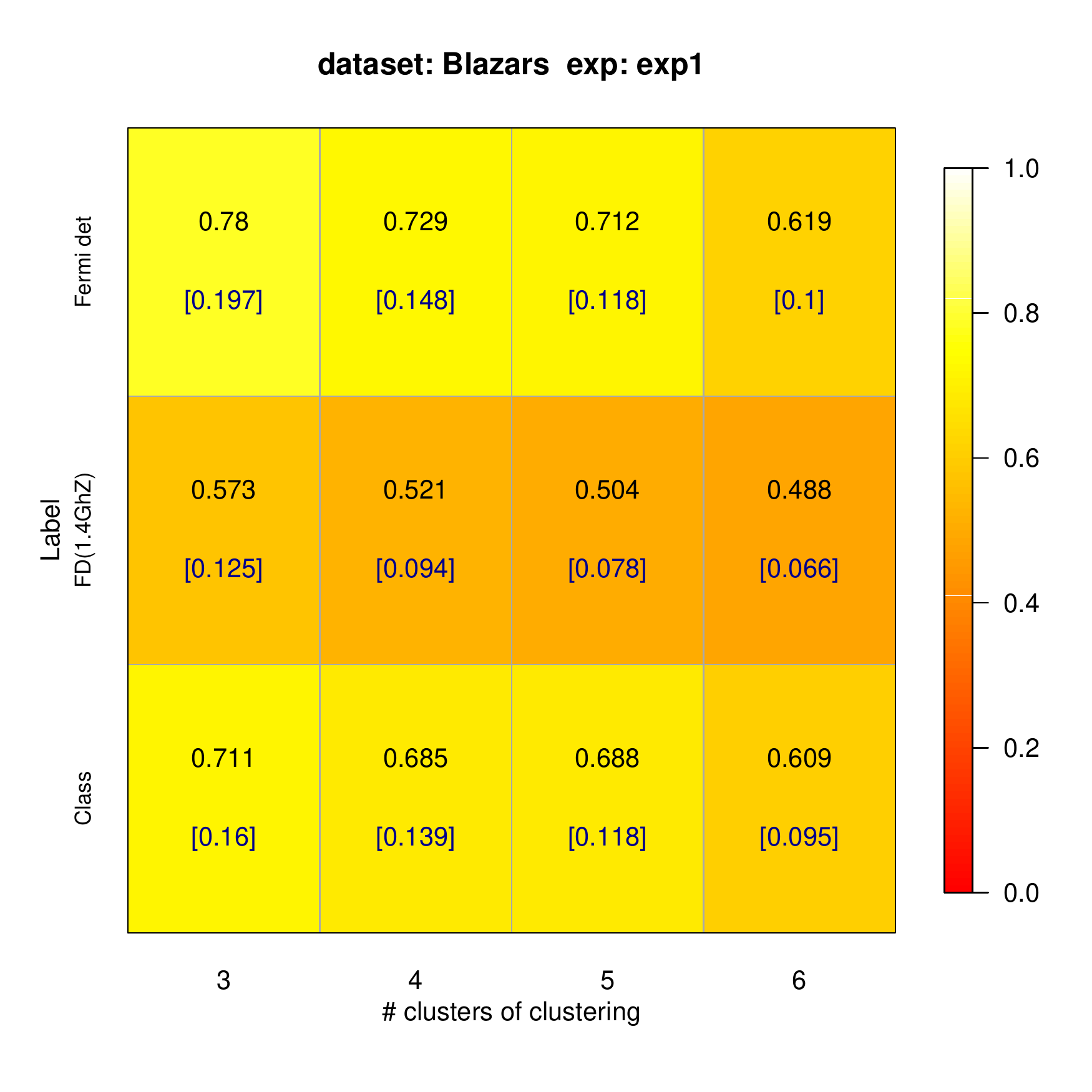}	&
\includegraphics[width=0.33\linewidth]{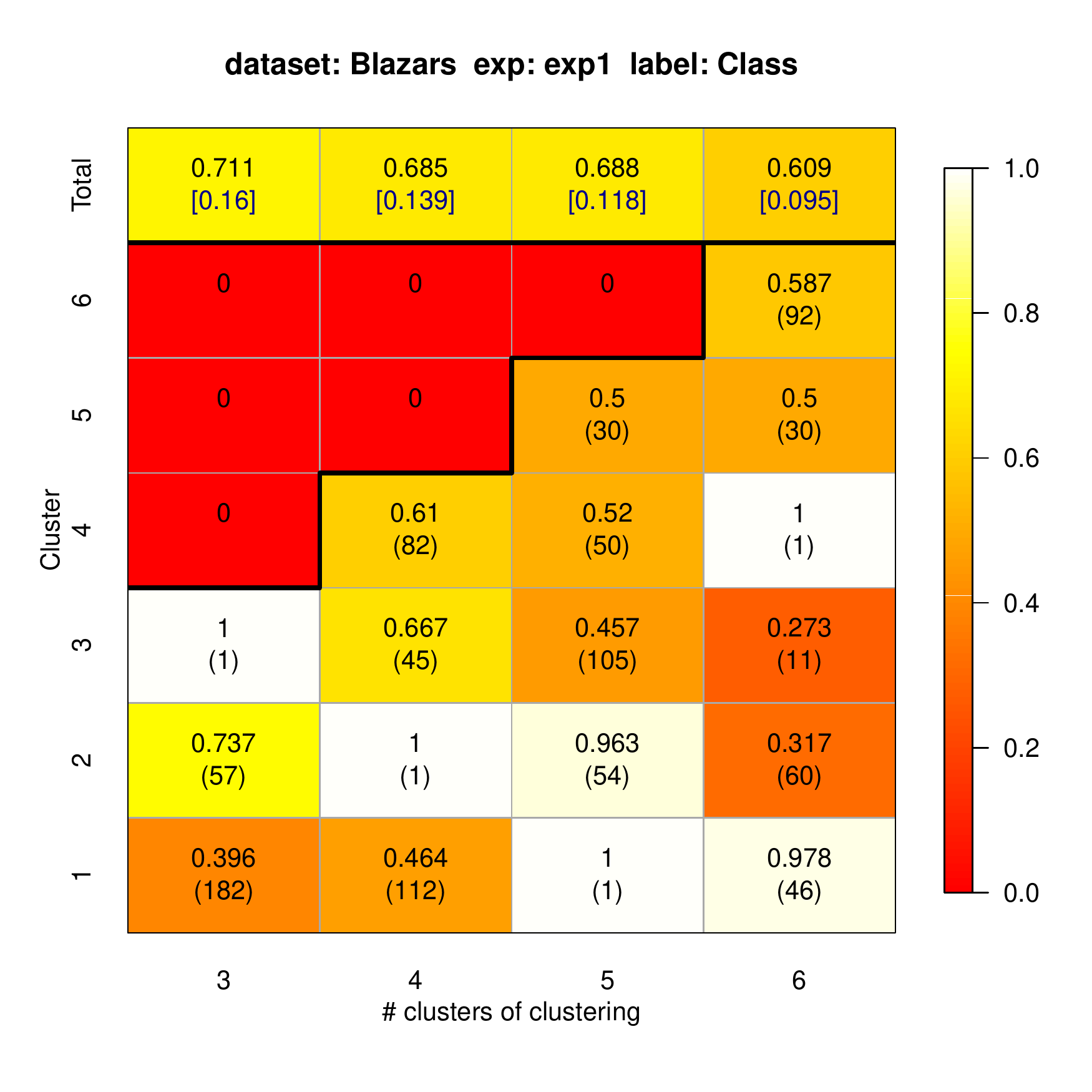}	&
\includegraphics[width=0.33\linewidth]{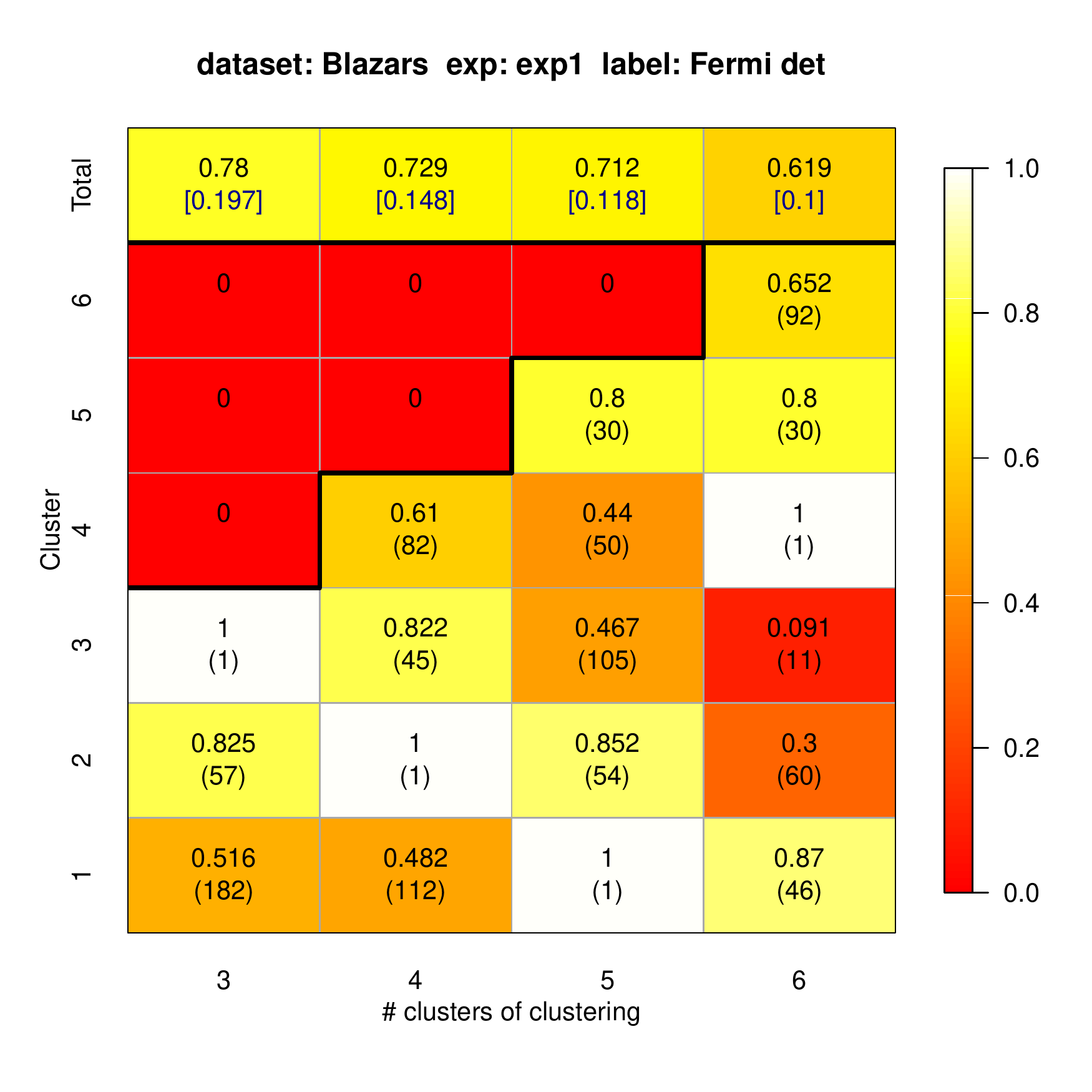}	\\
\includegraphics[width=0.33\linewidth]{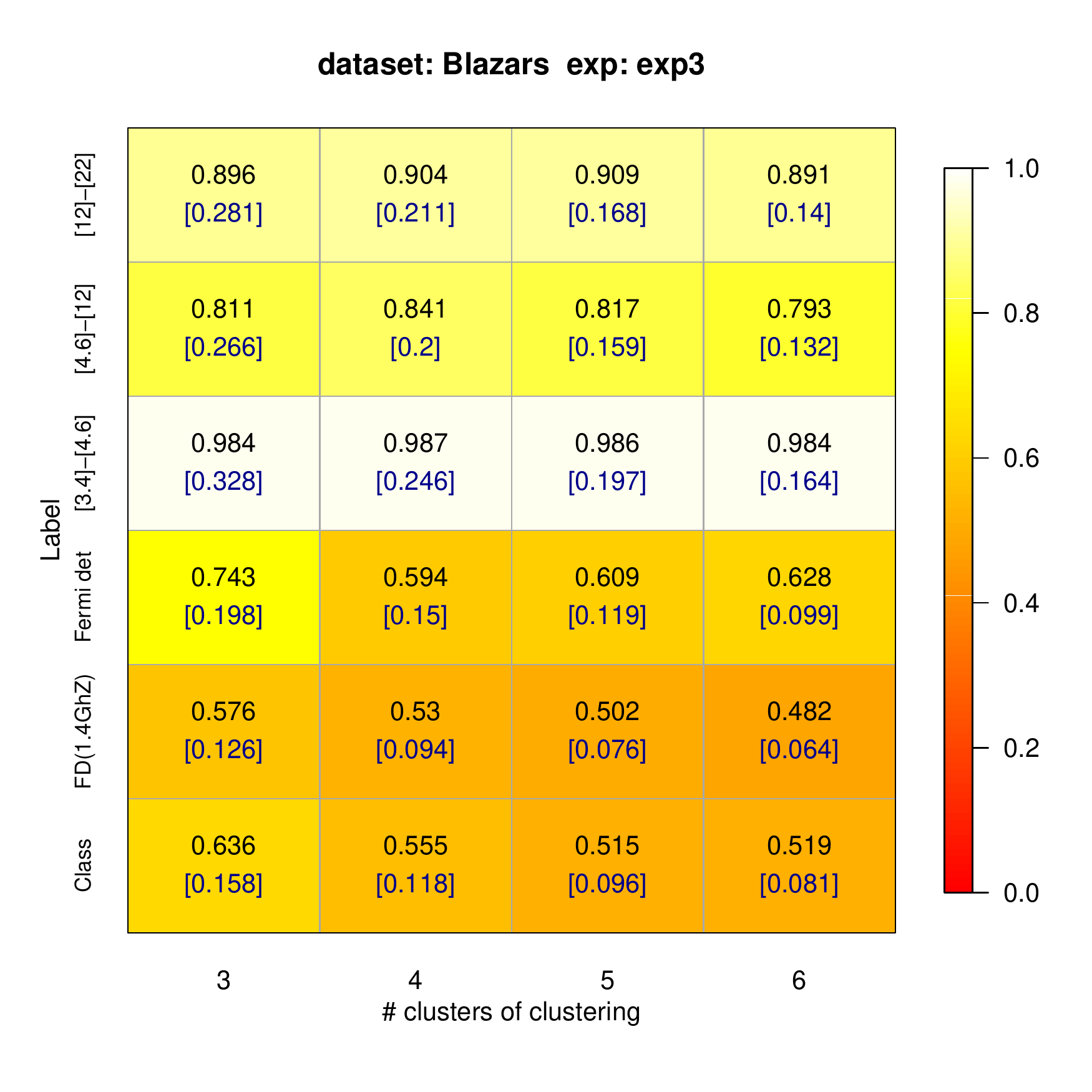}	&
\includegraphics[width=0.33\linewidth]{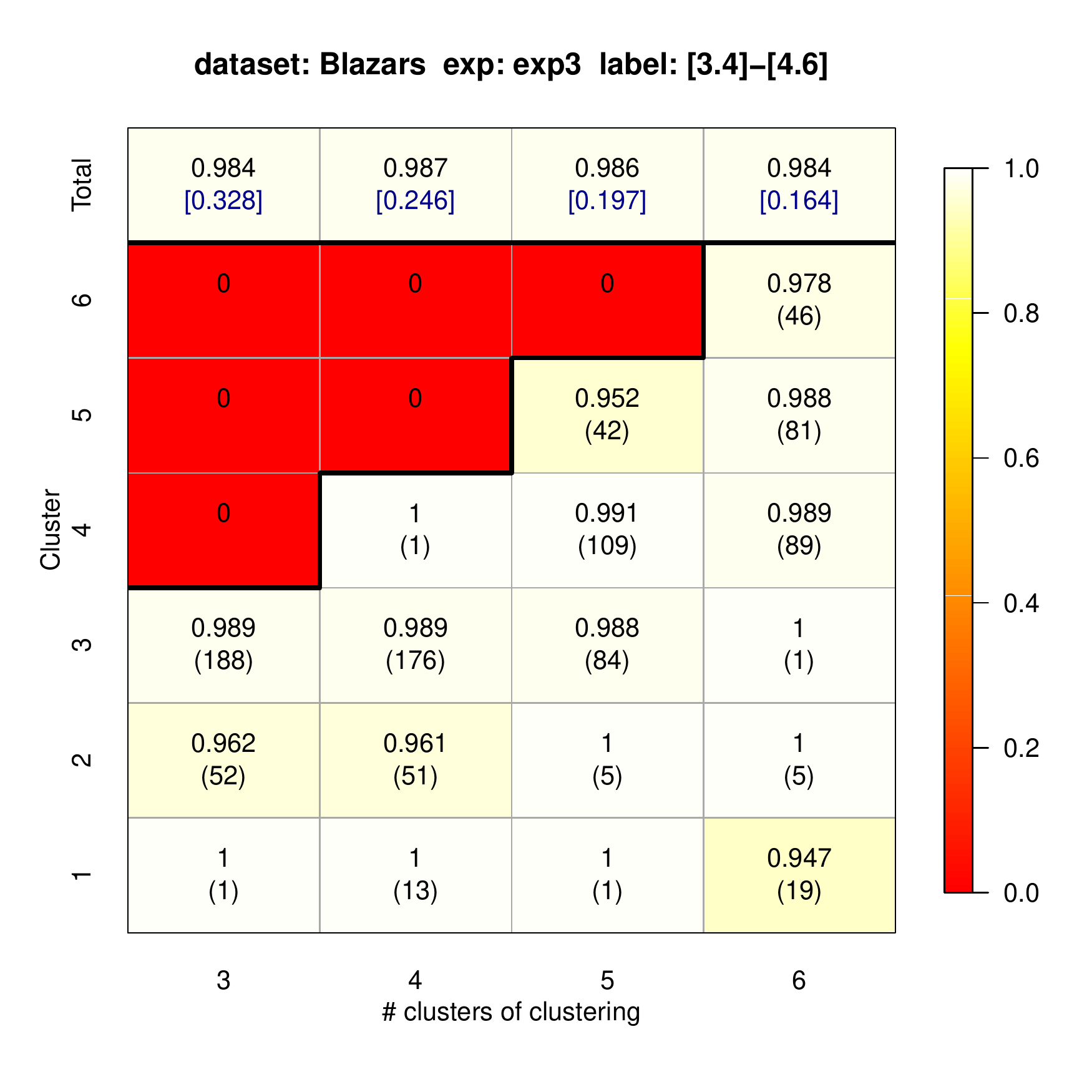}	&
\includegraphics[width=0.33\linewidth]{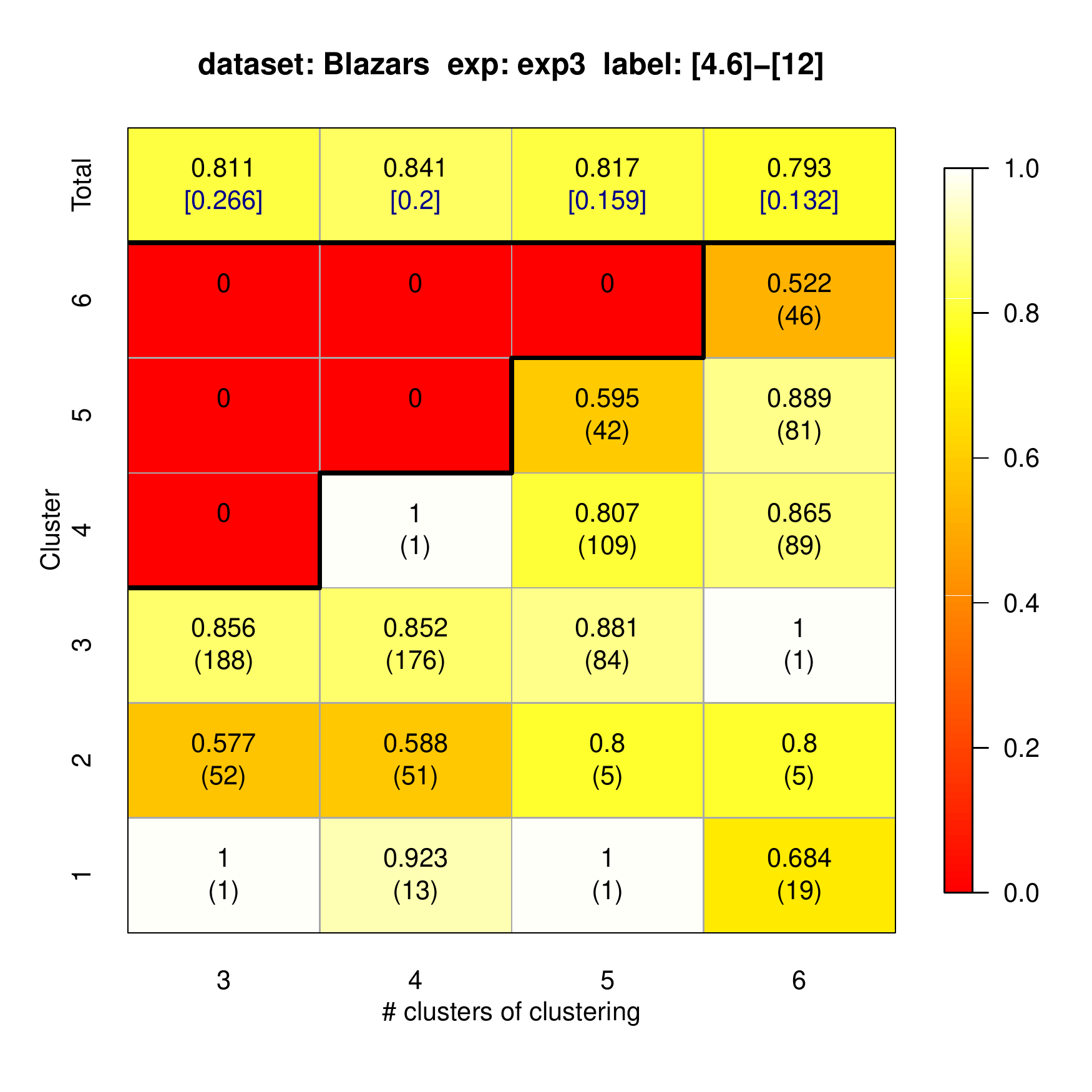}	\\
\end{tabular}
\end{center}
\caption{Results of the application of the CLaSPS method to the datasets based on the Blazars sample (for 
details see Sec.~\ref{sec:experiment2}). In the upper row, plots from the Exp.1 are shown. From left to right: 
total \emph{scores} distributions for clustering obtained with the K-means method; cluster \emph{scores} 
distributions for the spectral class of the blazars used as \emph{label}; cluster \emph{scores} distributions 
for the $\gamma$-ray detection flag used as \emph{label}. In the lower row, plots from the results of Exp.3 
are shown. From left to right: 
total \emph{score} distributions for clusterings created by the K-means method; cluster \emph{scores} 
distribution for the $[3.4]-[4.6]$ WISE color used as \emph{label}; cluster \emph{scores} distribution for the 
$[4.6]-[12]$ WISE color used as \emph{label}.}  
\label{fig:Blazarsexps}
\end{figure*}

The results of the application of CLaSPS to the three experiments based on the Blazars sample
and described in Table~\ref{tab:featureslabels2} are shown in the plots in Figure~\ref{fig:Blazarsexps}.

The main conclusion that can be drawn from the results of the first experiment (plots in the 
upper row in Figure~\ref{fig:Blazarsexps}),
is that the distribution of blazars in the optical+near-infrared+mid-infrared colors \emph{feature} 
space, consistently throughout the distinct clustering methods, strongly correlates with the 
spectral classification of the Blazars. The correlation is noticeable, based on the large values 
of the total normal and weighted \emph{scores} evaluated using the blazars spectral 
class as \emph{label} for all clustering (upper-mid plot in Figure~\ref{fig:Blazarsexps}). 
In particular, the largest total normal and weighted \emph{score} values for this
\emph{label} are both obtained for the clustering with total three clusters and produced with 
the K-means algorithm (upper-mid plot in Figure~\ref{fig:Blazarsexps}).  

\begin{figure*}
\begin{center}
\begin{tabular}{cc}
\includegraphics[width=0.5\linewidth]{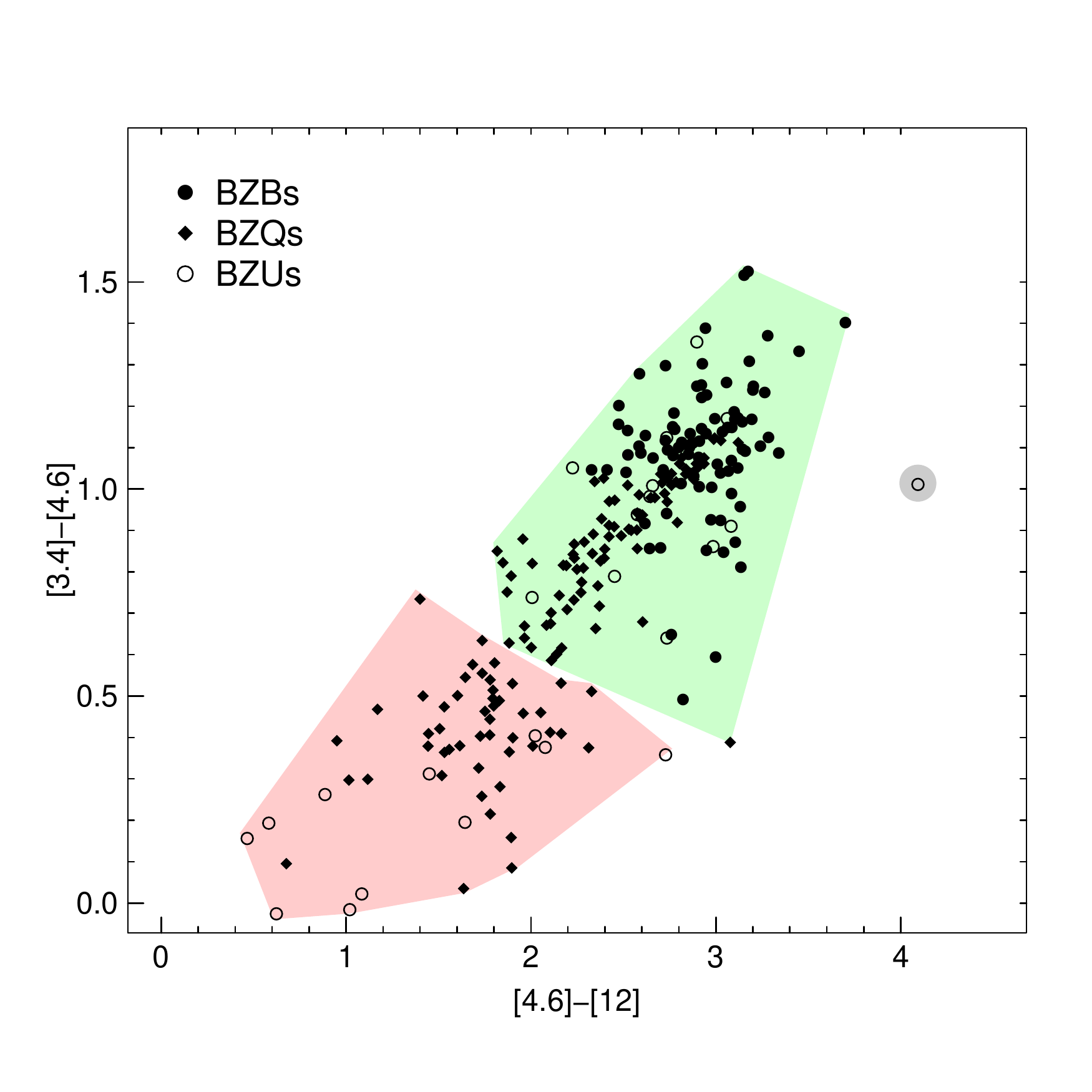} &
\includegraphics[width=0.5\linewidth]{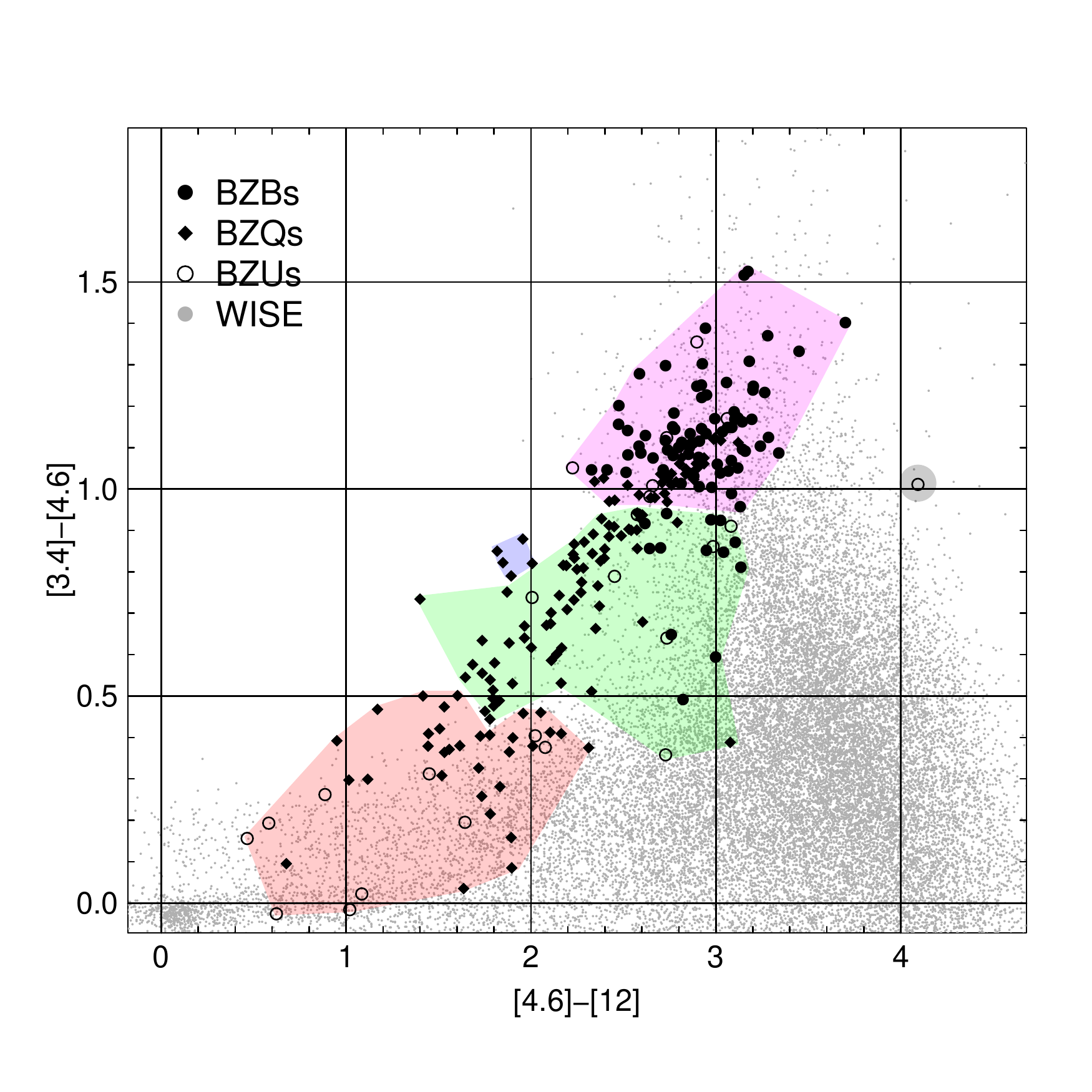}	\\
\end{tabular}
\end{center}
\caption{Left plot: projection of the distribution 
in the multi-dimensional \emph{feature} space of the Blazars sample used in the first experiment 
(Exp.~1 in Table~\ref{tab:featureslabels2}) onto the $[4.6]\!-\![12]$ vs $[3.4]\!-\![4.6]$ $\mu$m 
WISE color-color plane. Blazars 
of different spectral classes (BZB, BZQ and BZU) according to the BZCat are plotted with different 
symbols. 
The shaded regions correspond to the projections of the three clusters of the clustering
with three total clusters obtained with the K-means algorithm using the spectral class as
\emph{label} (upper-mid plot of Figure~\ref{fig:Blazarsexps}). The interpretation of the 
clustering is discussed in Sec.~\ref{subsec:application2}. Right plot: plot of the
distribution of the Blazars sample used in experiment three (Exp. 3) in the 
WISE color-color plane generated by the colors $[4.6]\!-\![12]$ and $[3.4]\!-\![4.6]$, used 
as \emph{labels}. The horizontal and vertical black lines
represent the edges of the \emph{label} bins used in the experiment three for the $[3.4]\!-\![4.6]$ 
and $[4.6]\!-\![12]$ WISE colors respectively (the distribution of \emph{scores} associated with 
these two \emph{labels} are shown in the lower central and left plots in Figure~\ref{fig:Blazarsexps}). 
The background grey dots are shown for reference and correspond to 453420 WISE 
generic sources detected at high Galactic latitude. This plot has been adapted from a 
similar plot in~\citep{dabrusco2012}.}  
\label{fig:Blazars_WISEcolors}
\end{figure*}

In order to verify whether there is a smaller subset of \emph{features} responsible for the  
correlation, a PCA was carried out on the \emph{feature} 
distribution of the two clusters with more than one member, of the 
clustering with three total clusters produced by the K-means with the (BZB, BZQ, BZU) spectral 
classes used as \emph{labels}. 
This analysis has shown that the spectral classification of the sources of the Blazars sample 
correlates very strongly with the mid-infrared colors from WISE.
The projection of the \emph{feature} space distribution of the sources in the Blazars sample 
onto the WISE $[4.6]-[12]$ vs $[3.4]-[4.6]$ color-color space is shown in the left plot in 
Figure~\ref{fig:Blazars_WISEcolors}. In this plot, the sources are plotted with different symbols
according to the value of the \emph{label}, the spectral class (BZB, BZQ or BZU), and the three
regions occupied by the projections of the three clusters of the clustering with total three clusters 
are represented as shaded colored areas.
This finding has been discussed in detail in~\citep{massaro2011}, 
where an explanation of the new correlation has been proposed, in terms of the 
currently accepted emission mechanisms of blazars. 

The same correlation between the distribution of \emph{features} of the clusterings 
and the spectral classification has been observed in the second experiment.
In this case, the dataset used contains only blazars from the ROMA-BZCAT catalog which 
have been associated with $\gamma$-ray sources in the 2FLG~\citep{abdo2011}. Similarly at 
what found for the second experiment (Exp.~2 in Table~\ref{tab:featureslabels2}), 
the correlation can be almost entirely ascribed to the peculiar WISE colors distribution of the 
blazars. 

An even stronger correlation has been observed in the third experiment. 
The third experiment has involved the same dataset used in the first experiment, comprising blazars
from the BZCat with optical and near-infrared colors as \emph{features} and 
the WISE infrared colors $[3.4]-[4.6]$, $[4.6]\!-\![12]$ and $[12]\!-\![22]$ as \emph{labels} instead. 
The three WISE colors, used as \emph{labels}, have been binned as shown in 
Table~\ref{tab:featureslabels2}. 

The lower central and left plots in Figure~\ref{fig:Blazarsexps} show the \emph{scores} values
distributions for the clusterings produced by the K-means 
method on the third experiment (Exp.~3 in Table~\ref{tab:featureslabels2}) calculated 
using as \emph{labels} the two colors 
$[3.4]\!-\![4.6]$ and $[4.6]\!-\![12]$ from WISE. The large values of the \emph{scores} for all
clusterings indicate that
a strong correlation exists between the distribution of sources in the clusterings and 
their mid-IR colors. Based on the binning used in this experiment for these two \emph{labels} 
(see Table~\ref{tab:featureslabels2}), this result suggests that the distribution of 
blazars in the mid-infrared colors is peculiar, as most blazars are contained in a 
narrow region of the mid-IR WISE colors space.  

This fact is evident in the right plot of Figure~\ref{fig:Blazars_WISEcolors}, where 
the projection onto the WISE $[3.4]\!-\![4.6]$ vs $[4.6]\!-\![12]$ color-color plane of the distribution of 
the blazars in the 
\emph{feature} space of the third experiment (Exp.~3 in Table~\ref{tab:featureslabels2}) 
is shown. The regions occupied by the 
projections of the clusters of the clustering with five total clusters produced by the 
K-means algorithm (see lower central and left plots in Figure~\ref{fig:Blazarsexps}) are 
represented by shaded colored areas. The symbols of the points, 
as in the left plot in Figure~\ref{fig:Blazars_WISEcolors}, reflect the spectral 
classification. The horizontal and vertical black lines correspond to the bin limits of the 
bins used for the $[3.4]\!-\![4.6]$ and $[4.6]\!-\![12]$ WISE colors used as \emph{labels} (see
Table~\ref{tab:featureslabels2} for more details about the experiment). The clusters in this plot
are associated with the values of the total and cluster \emph{scores} shown in the column
corresponding to the clustering with total five clusters in both the lower central and left plots
in Figure~\ref{fig:Blazarsexps}, for the two WISE colors respectively. The red, green and 
magenta large regions correspond to the second, third and fifth clusters of the clustering 
respectively, while the small
five-members group corresponds to the first cluster and the single-source cluster is the fourth
cluster in the two plots of Figure~\ref{fig:Blazarsexps}. Multiple generic sources from the 
WISE catalog, drawn from 
a region of the sky at high galactic latitude, are plotted as small gray points to show that
the \emph{locus} occupied by the blazars is clearly separated from the high-density regions
of the overall distribution of WISE sources in the same color-color plane. 

In this case, the application of CLaSPS has helped to determine a previously unknown 
correlation between 
a class of astronomical sources and a small number of \emph{features}, as the clusters
of all clusterings follow a narrow \emph{locus} in the WISE $[3.4]-[4.6]$ vs $[4.6]-[12]$ 
color-color plane, occupied by the whole sample of blazars. While this pattern is clearly 
visible also in the low-dimensional two/three dimensional distribution of the blazars WISE 
colors, it has been discovered during the exploration of the multi-dimensional \emph{feature} 
space generated by the multi-wavelength color of BZCat blazars. This is an example of
low-dimensional pattern that had gone, so far, unnoticed and that a general method for the
determination of correlations in complex \emph{feature} spaces, like CLaSPS, has helped
to single out and characterize further. 

A thorough investigation of the spectral mid-infrared and $\gamma$-ray 
properties of this sample of blazars that result in this peculiar pattern, has been 
presented in~\citep{dabrusco2012}. Some of the authors of this paper have also 
developed a method, based on the mid-infrared properties of BZCat blazars
discussed above~\citep{massaro2012}, for the selection of blazars candidates from 
WISE photometric data.

\section{Summary and future developments}
\label{sec:results}

In this paper we have presented CLaSPS, a new method for the determination of correlations in complex 
astronomical datasets, based on KD techniques for unsupervised 
clustering supplemented by the use of external information to label and characterize the content of the 
clusters (Sec.~\ref{sec:method}).
We have introduced the \emph{score} (Sec.~\ref{subsec:scores}) and shown the reliability of the 
\emph{score} as a measure of the degree of correlation among the membership distribution
of sources in a clustering and the distribution of a quantitative or categorial \emph{label} in distinct classes,
using simulated clusterings (Sec.~\ref{subsec:choice}). 

We have also discussed the applications of CLaSPS to two different 
samples composed of extragalactic sources with multi-wavelength photometry used as \emph{features}: the
first dataset, CSC+ (Sec.~\ref{sec:experiment1}), is composed of spectroscopically confirmed 
quasars from the SDSS DR8 
with multi-wavelength observations in the near-infrared, optical and ultraviolet, and detected 
(or with reliable upper limits) in the Chandra X-ray CSC catalog; the second dataset 
(Sec.~\ref{sec:experiment2}) is composed of 
optically confirmed blazars with mid-infrared, near-infrared and optical observations, complemented, 
for a subset of the sources, by $\gamma$-ray data from the 2FGL. 

The main result of the application of CLaSPS to the CSC+ dataset has been the confirmation of a well known
correlation~\citep[see, for example][]{lusso2010} between the near-ultraviolet/blue optical luminosity 
of optically selected radio-quiet quasars and the spectral index $\alpha_{\mathrm{OX}}$ 
(Sec.~\ref{subsec:application1}) in a subset of the highly 
inhomogeneous CSC+ sample. CLaSPS has narrowed the CSC+ sample to three specific clusters 
that show significant correlation between the $L_{\mathrm{opt}}(2500\AA)$ mono-chromatic luminosity 
and the $\alpha_{\mathrm{OX}}$ spectral index, based on the clustering of the CSC+ sample in the 
\emph{feature} space generated by the near-infrared, optical and ultraviolet photometric data. Further 
analysis of the results have shown that the correlation for the subset of sources contained in the correlated
clusters is driven by the values of the $nuv-u$ color, as an indicator of the presence of the ``big-blue-bump"
component in the SEDs of the sources.

In the case of the experiments performed on the blazars sample, CLaSPS has revealed
an unknown correlations between the spectral classification of the blazars 
in BZQs, BZBs and BZUs (Sec.~\ref{subsec:dataset2}) and their distribution 
in the \emph{feature} space generated by mid-infrared, near-infrared and optical colors. 
Further investigation has shown that the correlation is almost entirely attributable to the peculiar 
pattern followed by BZCat and $\gamma$-ray detected blazars follow in the WISE mid-infrared 
color space (Sec.~\ref{subsec:application2}). 
The implications of this pattern on the modeling of blazars emission mechanism and 
a novel method for the selection of candidate blazars from mid-infrared
survey photometric data based on such pattern have been investigated in 
other works by some of the authors~\citep{massaro2011,dabrusco2012,massaro2012}. 

While in this paper we have described applications of CLaSPS to inhomogeneous samples obtained 
by federating data from general purpose large area surveys, we plan to apply the method to large 
homogeneous samples of 
extragalactic sources, like the Chandra-{\it COSMOS} dataset~\citep{elvis2009,civano2012}.

CLaSPS selects the optimal clustering based on the \emph{scores}, a measure of the correlation 
between the clustering membership and a given partition of one external observable used as \emph{label}. 
For this reason, as discussed in Sec.~\ref{subsec:comparison}, 
CLaSPS differentiate itself from ``cluster ensembles'' techniques. Nonetheless,
three different aspects of the current CLaSPS method could be improved by the application of 
cluster ensembles techniques: 1) the limited number of 
clustering techniques used may bias the exploration of the clusterings towards 
particular aspects of the \emph{feature} distribution of the dataset considered. Moreover,
CLaSPS does not to take into account the properties and, potentially, weaknesses of each distinct
clustering techniques; 2) the choice of the optimal clustering is based on a single \emph{label} at the time.
Correlations between a given set of clusterings and multiple \emph{labels} cannot be captured 
by CLaSPS, but are left to the interpretation of multiple distinct \emph{label} 
experiments; 3) the choice of the optimal clusterings in CLaSPS is based on a single ``view" of the 
dataset, i.e. on clusterings obtained using a single set of sources and/or \emph{features}.

The first point could be easily addressed by widening the portfolio of clustering methods used by CLaSPS. 
Then, cluster ensembles methods could be applied to subsets of clusterings (grouped 
by total number of clusters or by type of clustering method) to determine the 
``consensus clustering'' of each subset of clusterings. The \emph{scores} would then be evaluated 
on the set of consensus clusterings determined in this way. The second point could be 
similarly addressed by searching for the ``consensus clusterings'' of the set of optimal clusterings 
selected through the \emph{scores} values for different \emph{labels}. 
 
The third point is particularly important for astronomy, because most astronomical datasets 
present different number of \emph{features} available for different members of the dataset.
In its current implementation, CLaSPS can be applied only to clusterings obtained with a fixed given
subset of sources and \emph{features}. CLaSPS, in this scenario, can be applied separately to distinct 
groups of sources in the dataset with a set of common \emph{features}. 
In order to overcome this limitation, distinct sets of clusterings could be obtained for different 
``views'' of the dataset, i.e. different subsets of the datasets with the same set of \emph{features} 
available. Then, the multiple clusterings obtained with on the different views of the dataset with
different clustering techniques could be consolidated into a single set of clusterings through the 
application of cluster ensembles technique on the groups of clustering obtained with the same
clustering technique on distinct views of the dataset. This approach is similar to 
``\emph{features} distributed clustering'' and ``object distributed clustering" scenarios typical 
of practical application of clustering ensemble \citep{strehl2003}. 

A further improvement to the CLaSPS method is related to the choice of the classes of the 
\emph{labels}. In the frequent case of quantitative continuous \emph{labels}, the choice of the binning 
is crucial for the evaluation of the \emph{scores} and, in turn, for the determination of the correlations 
among \emph{features} and \emph{labels}, if any. While the astronomer deciding the binning of the 
\emph{labels} on the basis of \emph{a priori} knowledge of the specific topic considered is a viable 
option for most cases where the astronomer tries to generalize an already known correlation or 
a generic problem (e.g. the characterization of astronomical sources based on their
photometric parameters for this paper) is investigated, 
this can be a limitation to the generality of the method when the aim of the experiments is 
a ``blind" exploration of multi-dimensional astronomical datasets.
In order to improve this aspect of the CLaSPS method, we are exploring the possibility of 
complementing the astronomer's definition of classes of \emph{labels} with 
spontaneous classes that can be determined from the intrinsic distribution of the 
\emph{labels} themselves by the application of non-parametric KD techniques.

\acknowledgements

R. D'Abrusco acknowledges the financial support of the US Virtual Astronomical Observatory, 
which is sponsored by the National Science Foundation and the National Aeronautics and Space 
Administration. We acknowledge partial support by NASA contract NAS-39073 (CXC).
The CLaSPS method is implemented in \emph{R}~\citep{ihaka1996}, an 
open-source free statistical environment developed under the GNU GPL 
(\underline{http://www.r-project.org/}).
TOPCAT and STILTS 
(\underline{http://www.star.bris.ac.uk/$\sim$mbt/topcat/})~\citep{taylor2005} were extensively 
used for the preparation and manipulation of the tabular data in this work.


\clearpage

\end{document}